\documentclass[printer,traditabstract]{aa}
\usepackage{natbib}
\bibpunct{(}{)}{;}{a}{}{,}

\usepackage{setspace,ulem}
\usepackage{psfig,epsfig,amssymb,alltt}
\usepackage{latexsym}
\usepackage{indentfirst}
\usepackage[english]{babel}
\usepackage{graphicx}
\usepackage{epstopdf}
 
\usepackage{chemarrow}
\usepackage{psfig}
\usepackage{vmargin}
\usepackage{fancybox}
\usepackage{fancyheadings}
\usepackage{algorithmic}
\usepackage{algorithm}
\usepackage{theorem}
\usepackage{amssymb,amsmath}
 \usepackage{longtable,multirow,enumerate,threeparttable}
 \usepackage{amsfonts,upgreek,bbm,mathrsfs}

\usepackage{url}
\usepackage[applemac]{inputenc}
\usepackage[T1]{fontenc}
\vspace{1cm}

\newtheorem{lemme}{Lemma}

\date\today
\title{Poisson Denoising on the Sphere: Application to the Fermi Gamma Ray Space Telescope}
\author{J. Schmitt \inst{1}
\and J.L. Starck \inst{1}
\and J.M. Casandjian \inst{1}
\and J. Fadili \inst{2}
\and I. Grenier \inst{1}
}

\authorrunning{Schmitt et al}
\titlerunning{Poisson Denoising on the Sphere}

\institute{
\inst{1} CEA, Laboratoire AIM, CEA/DSM-CNRS-Universit\'e Paris Diderot,   
CEA, IRFU, Service d'Astrophysique,  Centre de Saclay,  F-91191 Gif-Sur-Yvette cedex, France \\
\inst{2} GREYC CNRS-ENSICAEN-Universit\'e de Caen, 6, Bd du Mar\'echal Juin, 14050 Caen Cedex, France}

\begin{document}

\abstract{
The Large Area Telescope (LAT), the main instrument of the Fermi Gamma-Ray Space Telescope, detects high energy gamma rays with energies from 20 MeV to more than 300 GeV. The two main scientific objectives, the study of the Milky Way diffuse background and the detection of point sources, are complicated by the lack of photons. That is why we need a powerful Poisson noise removal method on the sphere which is efficient on low count Poisson data. This paper presents a new multiscale decomposition on the sphere for data with Poisson noise, called Multi-Scale Variance Stabilizing Transform on the Sphere (MS-VSTS). This method is based on a Variance Stabilizing Transform (VST), a transform which aims to stabilize a Poisson data set such that each stabilized sample has a quasi constant variance. In addition, for the VST used in the method, the transformed data are asymptotically Gaussian. MS-VSTS consists of decomposing the data into a sparse multi-scale dictionary like wavelets or curvelets, and then applying a VST on the coefficients in order to get almost Gaussian stabilized coefficients. In this work, we use the Isotropic Undecimated Wavelet Transform (IUWT) and the Curvelet Transform as spherical multi-scale transforms. Then, binary hypothesis testing is carried out to detect significant coefficients, and the denoised image is reconstructed with an iterative algorithm based on Hybrid Steepest Descent (HSD). To detect point sources, we have to extract the Galactic diffuse background: an extension of the method to background separation is then proposed. In contrary, to study the Milky Way diffuse background, we remove point sources with a binary mask. The gaps have to be interpolated: an extension to inpainting is then proposed. The method, applied on simulated Fermi LAT data, proves to be adaptive, fast and easy to implement.
}
\keywords{methods: Data Analysis --  techniques: Image Processing}

\maketitle

\section{Introduction}

The Fermi Gamma-ray Space Telescope, which was launched by NASA in june 2008, is a powerful space observatory which studies the high-energy gamma-ray sky \citep{Fermi}. 
Fermi's main instrument, the Large Area Telescope (LAT), detects photons in an energy range between 20 MeV to greater than 300 GeV. The LAT is much more sensitive than its predecessor, the EGRET telescope on the Compton Gamma Ray Observatory, and is expected to find several thousand gamma ray sources, which is an order of magnitude more than its predecessor EGRET~\citep{egret:hartman99}.

Even with its $\mathbf{ \sim 1 m^2}$ effective area, the number of photons detected by the LAT outside the Galactic plane and away from intense sources is expected to be low. Consequently, the spherical photon count images obtained by Fermi are degraded by the fluctuations on the number of detected photons. The basic photon-imaging model assumes that the number of detected photons at each pixel location is Poisson distributed. More specifically, the image is considered as a realization of an inhomogeneous Poisson process. This quantum noise makes the source detection more difficult, consequently it is  better to have an efficient denoising method for spherical Poisson data.

Several techniques have been proposed in the literature to estimate Poisson intensity in 2D. A major class of methods adopt a multiscale bayesian framework specifically tailored for Poisson data~\citep{KolaczykNowak2000}, independently initiated by \citet{Timmerman} and \citet{Kolaczyk}. \citet{Lefkimmiatis} proposed an improved bayesian framework for analyzing Poisson processes, based on a multiscale representation of the Poisson process in which the ratios of the underlying Poisson intensities in adjacent scales are modeled as mixtures of conjugate parametric distributions. Another approach includes preprocessing the count data by a variance stabilizing transform (VST) such as the \citet{Anscombe} and the \citet{Fisz} transforms, applied respectively in the spatial~\citep{Donoho} or in the wavelet domain~\citep{Fryzlewicz}. The transform reforms the data so that the noise approximately becomes Gaussian with a constant variance. Standard techniques for independant identically distributed Gaussian noise are then used for denoising. \citet{Zhang} proposed a powerful method called Multi-Scale Variance Stabilizing Tranform (MS-VST). It consists in combining a VST with a multiscale transform (wavelets, ridgelets or curvelets), yielding asymptotically normally distributed coefficients with known variances. The choice of the multi-scale method depends on the morphology of the data. Wavelets represent more efficiently regular structures and isotropic singularities, whereas ridgelets are designed to represent global lines in an image, and curvelets represent efficiently curvilinear contours. Significant coefficients are then detected with binary hypothesis testing, and the final estimate is reconstructed with an iterative scheme. In \citet{Starck09:fermi3d}, it was shown that sources can be detected in 3D LAT data (2D+time or 2D+energy) using a specific 3D extension of the MS-VST.

There is, to our knowledge, no method for Poisson intensity estimation on spherical data. It is possible to decompose the spherical data into several 2D projections, denoise each projection and reconstitute the denoised spherical data, but the projection induces some caveats like visual artifacts on the borders or deformation of the sources.

In the scope of the Fermi mission, we have two main scientific objectives:
\begin{itemize}
  \item Detection of point sources to build the catalog of gamma ray sources,
  \item Study of the Milky Way diffuse background, which is due to interaction between cosmic rays and interstellar gas and radiation.
\end{itemize}

The first objective implies the extraction of the galactic diffuse background. Consequently, we want a method to suppress Poisson noise while extracting a model of the diffuse background. The second objective implies the suppression of the point sources: we want to apply a binary mask on the data (equal to $0$ on point sources, and to $1$ everywhere else) and to denoise the data while interpolating the missing part. Both objectives are linked: a better knowledge of the Milky Way diffuse background enables us to improve our background model, which leads to a better source detection, while the detected sources are masked to study the diffuse background.

The aim of this paper is to introduce a Poisson denoising method on the sphere called Multi-Scale Variance Stabilizing Transform on the Sphere (MS-VSTS) in order to denoise the Fermi photon count maps.
This method is based on the MS-VST~\citep{Zhang} and on recent on multi-scale transforms on the sphere \citep{starck2006,Abrial,starck:abrial08}.
Section~2 recalls the multiscale transforms on the sphere which are used in this paper, and 
 Gaussian denoising methods based on sparse representations. 
 Section~3 introduces the MS-VSTS. 
 Section~4 applies the MS-VSTS to spherical data restoration. Section~5 applies the MS-VSTS to inpainting. Section~6 applies the MS-VSTS to background extraction. Conclusions are drawn in Section~7. In this paper, all experiments were performed on HEALPix maps with $nside=128$~\citep{Gorski}, which corresponds to a good pixelisation choice for  the GLAST/FERMI resolution. The performance of the method is not dependent on the nside parameter. For a given data set, if nside is small, it just means that we don't want to investigate the finest scales. If nside is large, the number of counts per pixel will be very small, and we may not have enough statistics to get any information at the finest resolution levels. But it will not have any bad effect on the solution. Indeed,  the finest scales
will be smoothed, since our algorithm will not detect any significant wavelet coefficients in the finest scales. Hence, starting with a fine pixelisation (i.e. large nside), our method will provide a kind of automatic binning, by thresholding wavelets coefficients at scales and at spatial positions where the number of counts is not sufficient.

\section{Multi-scale analysis on the sphere}

New multi-scale transforms on the sphere were developed by \citet{starck2006}. These transforms can be inverted and  are easy to compute with the HEALPix pixellisation, and were used for denoising, deconvolution, morphological component analysis and impainting applications~\citep{Abrial}. In this paper, we use the Isotropic Undecimated Wavelet Transform (IUWT) and the Curvelet Transform.

\subsection{Multi-scale Transforms on the sphere}

\subsubsection{Isotropic Undecimated Wavelet Transform on the sphere}

The Isotropic Undecimated Wavelet Transform on the sphere (IUWT) is a wavelet transform on the sphere based on the spherical harmonics transform and with a very simple reconstruction algorithm. At scale $j$, we denote $a_j(\theta,\varphi)$ the scale coefficients, and $d_j(\theta,\varphi)$ the wavelet coefficients, with $\theta$ denoting the longitude and $\varphi$ the latitude. Given a scale coefficient $a_j$, the smooth coefficient $a_{j+1}$ is obtained by a convolution with a low pass filter $h_j$ : $a_{j+1} = a_j \ast h_j$. The wavelet coefficients are defined by the difference between two consecutive resolutions : $d_{j+1} = a_j - a_{j+1}$. A straightforward reconstruction is then given by:
\begin{equation}
\label{eq9}
a_0(\theta,\varphi) = a_J(\theta,\varphi) + \sum_{j=1}^{J}d_j(\theta,\varphi)
\end{equation}
Since this transform is redundant, the procedure for reconstructing an image from its coefficients is not unique and this can be profitably used to impose additional constraints on the synthesis functions (e.g. smoothness, positivity). A reconstruction algorithm based on a variety of filter banks is described in \citet{starck2006}.

\subsubsection{Curvelet Transform on the sphere}

The curvelet transform enables the directional analysis of an image in different scales. The data undergo an Isotropic Undecimated Wavelet Transform on the sphere. Each scale $j$ is then decomposed into smoothly overlapping blocks of side-length $B_j$ in such a way that the overlap between two vertically adjacent blocks is a rectangular array of size $B_j \times B_j /2$, using the HEALPix pixellisation. Finally, the ridgelet transform~\citep{CandesDonoho1999} is applied on each individual block. The method is best for the detection of anisotropic structures and smooth curves and edges of different lengths. More details can be found in  \citet{starck2006}.

\subsection{Application to gaussian denoising on the sphere}

Multiscale transforms on the sphere have been used successfully for Gaussian denoising via non-linear filtering or thresholding methods. Hard thresholding, for instance, consists \textbf{of} setting all insignificant coefficients (i.e. coefficients with an absolute value below a given threshold) to zero. In practice, we need to estimate the noise standard deviation $\sigma_j$ in each band $j$ and a coefficient $w_j$ is significant if $|w_j| > \kappa \sigma_j$, where $\kappa$ is a parameter typically chosen between $3$ and $5$. Denoting $\mathbf{Y}$ the noisy data and $HT_{\lambda}$ the thresholding operator, the filtered data $\mathbf{X}$ are obtained by:
\begin{equation}
\label{ }
\mathbf{X} = \mathbf{\Phi} HT_{\lambda} (\mathbf{\Phi}^{T}\mathbf{Y}),
\end{equation}
where $\mathbf{\Phi}^{T}$ is the multiscale transform (IUWT or curvelet) and $\mathbf{\Phi}$ is the multiscale reconstruction. $\lambda$ is a vector which has the size of the number of bands in the used multiscale transform. The thresholding operation thresholds all coefficients in band $j$ with the threshold $\lambda_j = \kappa \sigma_j$.

\section{Multi-Scale Variance Stabilizing Transform on the Sphere (MS-VSTS)}

\subsection{Principle of VST}

\subsubsection{VST of a Poisson process}

Given Poisson data $\mathbf{Y} := (Y_i)_i$, each sample $Y_i \sim \mathcal{P} (\lambda_i)$ has a variance $\text{Var}[Y_i] = \lambda_i$. Thus, the variance of $\mathbf{Y}$ is signal-dependant. The aim of a VST $\mathbf{ T}$ is to stabilize the data such that each coefficient of $\mathbf{ T}(\mathbf{Y})$ has an (asymptotically) constant variance, say $1$, irrespective of the value of $\lambda_i$. In addition, for the VST used in this study, $T(\mathbf{Y})$ is asymptotically normally distributed. Thus, the VST-transformed data are asymptotically stationary and gaussian.

The \citet{Anscombe} transform is a widely used VST which has a simple square-root form
\begin{equation}
\label{eq14}
\mathbf{ T}(Y):=2\sqrt{Y+3/8}.
\end{equation}
We can show that $\mathbf{ T}(Y)$ is asymptotically normal as the intensity increases.
\begin{equation}
\label{eq15}
\mathbf{ T}(Y)-2\sqrt{\lambda} \autorightarrow{$\mathcal{D}$}{$\lambda \rightarrow + \infty$} \mathcal{N}(0,1)
\end{equation}
It can be shown that the Anscombe VST requires a high underlying intensity to well stabilize the data (typically for $\lambda \geqslant 10$) \citep{Zhang}.

\subsubsection{VST of a filtered Poisson process}

Let $Z_j := \sum_i h[i] Y_{j-i}$ be the filtered process obtained by convolving $(Y_i)_i$ with a discrete filter $h$. We will use $Z$ to denote any of the $Z_j$'s. Let us define $\tau_k := \sum_i (h[i])^k$ for $k=1,2,\cdots$. In addition, we adopt a local homogeneity assumption stating that $\lambda_{j-i} = \lambda$ for all $i$ within the support of $h$.

We define the square-root transform $T$ as follows:
\begin{equation}
\label{eq16}
T(Z):=b\cdot \mathrm{sign}(Z+c) |Z+c|^{1/2},
\end{equation}
where $b$ is a normalizing factor. Lemma \ref{lemme1} proves that $T$ is a VST for a filtered Poisson process (with a nonzero-mean filter) in that $T(Y)$ is asymptotically normally distributed with a stabilized variance as $\lambda$ becomes large (see \citet{Zhang} for a proof).

\begin{lemme} \label{lemme1}
\emph{\textbf{(Square root as VST)}} If $\tau_1 \neq 0$, $\|h\|_2,\|h\|_3<\infty$, then we have : \\
\begin{equation} \label{eq17}
\begin{split}
\mathrm{sign}(Z+c)\sqrt{|Z+c|}-\mathrm{sign}(\tau_1)\sqrt{|\tau_1|\lambda} \\
 \autorightarrow{$\mathcal{D}$}{$\lambda \rightarrow + \infty$} \mathcal{N}\Big(0,\frac{\tau_2}{4|\tau_1|}\Big).
\end{split}
\end{equation}
\end{lemme}

\subsection{MS-VSTS}

The MS-VSTS consists in combining the square-root VST with a multi-scale transform.

\subsubsection{MS-VSTS + IUWT}

This section describes the MS-VSTS + IUWT, which is a combination of a square-root VST with the IUWT. The recursive scheme is:
\begin{equation}
\label{eq27}
\begin{split}
&\text{IUWT}\left\{\begin{array}{ccc}a_j  & = &  h_{j-1} \ast a_{j-1}  \\d_j  & = & a_{j-1}  - a_j  \end{array}\right. \\
 \Longrightarrow & \begin{split}\text{MS-VSTS} \\  \text{+ IUWT} \end{split}\left\{\begin{array}{ccc}a_j  & = &  h_{j-1} \ast a_{j-1} \\d_j  & = & T_{j-1}(a_{j-1}) - T_j(a_j) \end{array}\right. .
\end{split}
\end{equation}

In (\ref{eq27}), the filtering on $a_{j-1}$ can be rewritten as a filtering on $a_0 := \mathbf{Y}$, i.e., $a_j = h^{(j)} \ast a_0$, where $h^{(j)} = h_{j-1} \ast \cdots \ast h_{1} \ast h_0$ for $j \geqslant 1$ and $h^{(0)} = \delta$, where $\delta$ is the Dirac pulse ($\delta = 1$ on a single pixel and $0$ everywhere else). $T_j$ is the VST operator at scale $j$:
\begin{equation}
\label{eq28}
T_j(a_j) = b^{(j)} \mathrm{sign}(a_j+c^{(j)})\sqrt{|a_j + c^{(j)}|} .
\end{equation}
Let us define $\tau_k^{(j)}:=\sum_i (h^{(j)}[i])^k$. In~\citet{Zhang}, it has ben shown that, to have an optimal convergence rate for the VST, the constant $c^{(j)}$ associated to $h^{(j)}$ should be set to:
\begin{equation}
\label{eq29}
c^{(j)}:=\frac{7\tau_2^{(j)}}{8\tau_1^{(j)}} - \frac{\tau_3^{(j)}}{2\tau_2^{(j)}} .
\end{equation}
The MS-VSTS+IUWT procedure is directly invertible as we have:
\begin{equation}
\label{eq30}
a_0 (\theta,\varphi) = T_0^{-1} \Bigg[ T_J(a_J) + \sum_{j=1}^J d_j \Bigg] (\theta,\varphi).
\end{equation}
Setting $b^{(j)}:=\text{sgn}(\tau_1^{(j)})/\sqrt{|\tau_1^{(j)}|}$, if $\lambda$ is constant within the support of the filter.
$h^{(j)}$, then we have \citep{Zhang}:
\begin{equation}
\label{eq31}
\begin{split}
d_j(\theta,\varphi) \autorightarrow{$\mathcal{D}$}{$\lambda \rightarrow + \infty$} 
\mathcal{N} \Bigg( 0 , \frac{\tau_2^{(j-1)}}{4\tau_1^{(j-1)^2}} +\\ \frac{\tau_2^{(j)}}{4\tau_1^{(j)^2}} - \frac{\langle h^{(j-1)},h^{(j)} \rangle}{2\tau_1^{(j-1)}\tau_1^{(j)}} \Bigg) ,
\end{split}
\end{equation}
where $\langle . , . \rangle$ denotes inner product.

It means that the detail coefficients issued from locally homogeneous parts of the signal follow asymptotically a central normal distribution with an intensity-independant variance which relies solely on the filter $h$ and the current scale for a given filter $h$. Consequently, the stabilized variances and the constants $b^{(j)}$,$c^{(j)}$,$\tau_k^{(j)}$ can all be pre-computed.
Let us define $\sigma_{(j)}^2$ the stabilized variance at scale $j$ for a locally homogeneous part of the signal:
\begin{equation}
\label{eq32}
\sigma_{(j)}^2 = \frac{\tau_2^{(j-1)}}{4\tau_1^{(j-1)^2}} + \frac{\tau_2^{(j)}}{4\tau_1^{(j)^2}} - \frac{\langle h^{(j-1)},h^{(j)} \rangle}{2\tau_1^{(j-1)}\tau_1^{(j)}} .
\end{equation}

To compute the $\sigma_{(j)}$, $b^{(j)}$,$c^{(j)}$,$\tau_k^{(j)}$, we only have to know the filters $h^{(j)}$. We compute these filters thanks to the formula $a_j = h^{(j)} \ast a_0$, by applying the IUWT to a Dirac pulse $a_0 = \delta$. Then, the $h^{(j)}$ are the scaling coefficients of the IUWT. The $\sigma_{(j)}$ have been precomputed for a 6-scaled IUWT (Table~\ref{sigmaj}). 

\begin{table*}[!h]
  \centering
    \caption{Precomputed values of the variances $\sigma_j$ of the wavelet coefficients.
  }
  \begin{tabular}{|c|c|}
\hline
Wavelet scale $j$ & Value of $\sigma_j$ \\
\hline
  1 & 0.484704 \\
  2 & 0.0552595 \\
  3 & 0.0236458 \\
  4 & 0.0114056 \\
  5 & 0.00567026 \\
\hline
\end{tabular}

  \label{sigmaj}
\end{table*}

\begin{figure*}[htb]
\centering
\includegraphics[width=2.9in]{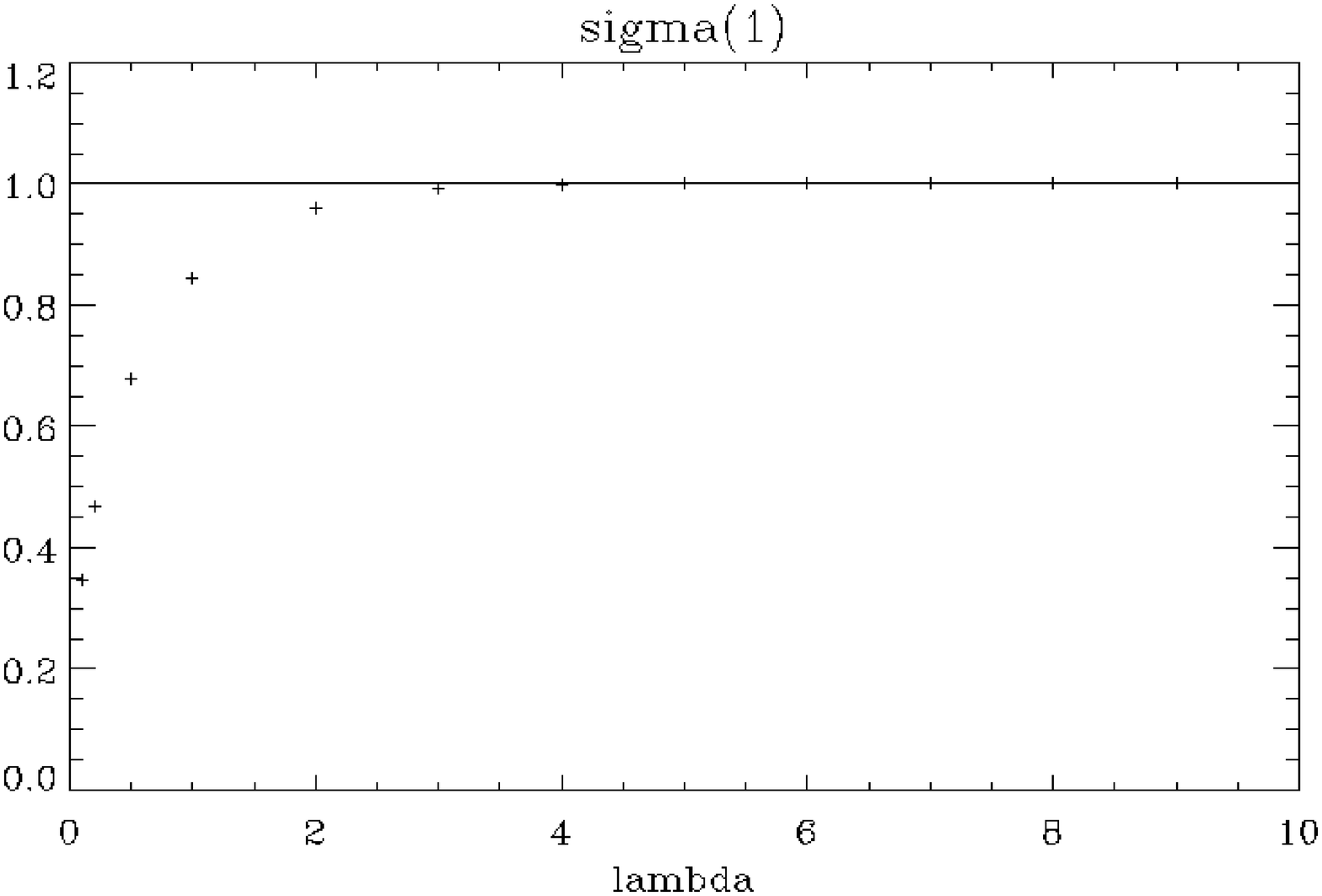}
\includegraphics[width=2.9in]{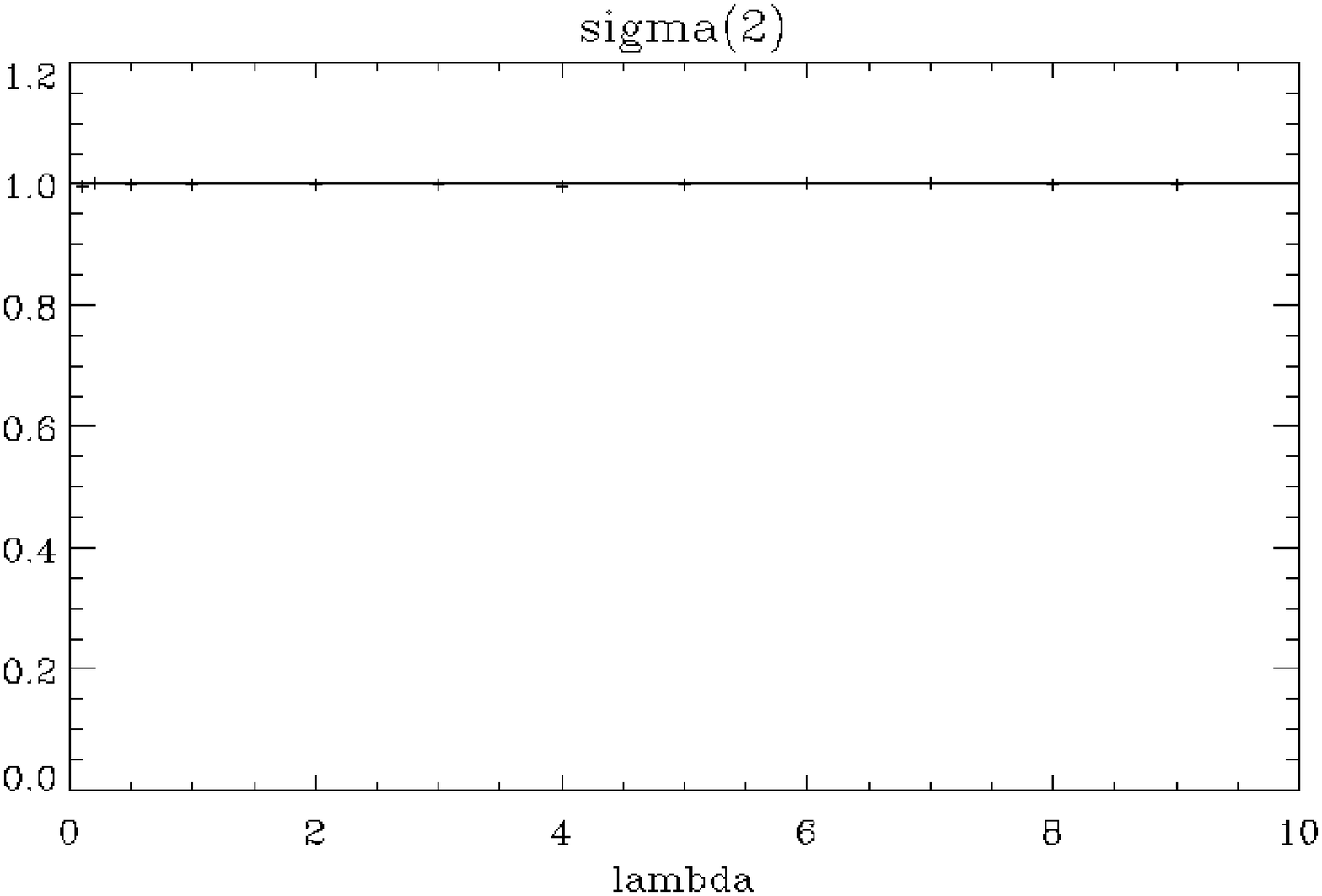}
\includegraphics[width=2.9in]{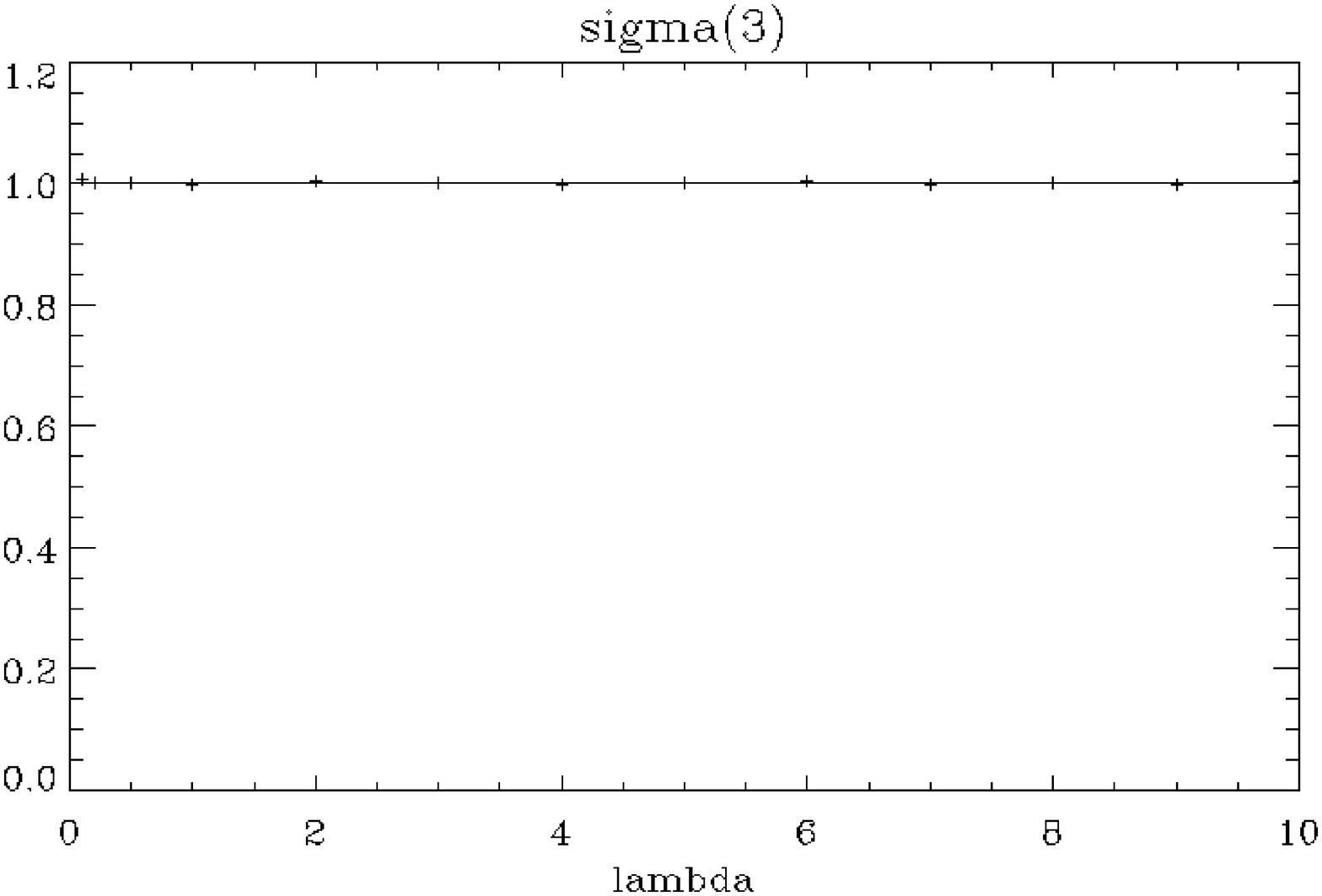}
\includegraphics[width=2.9in]{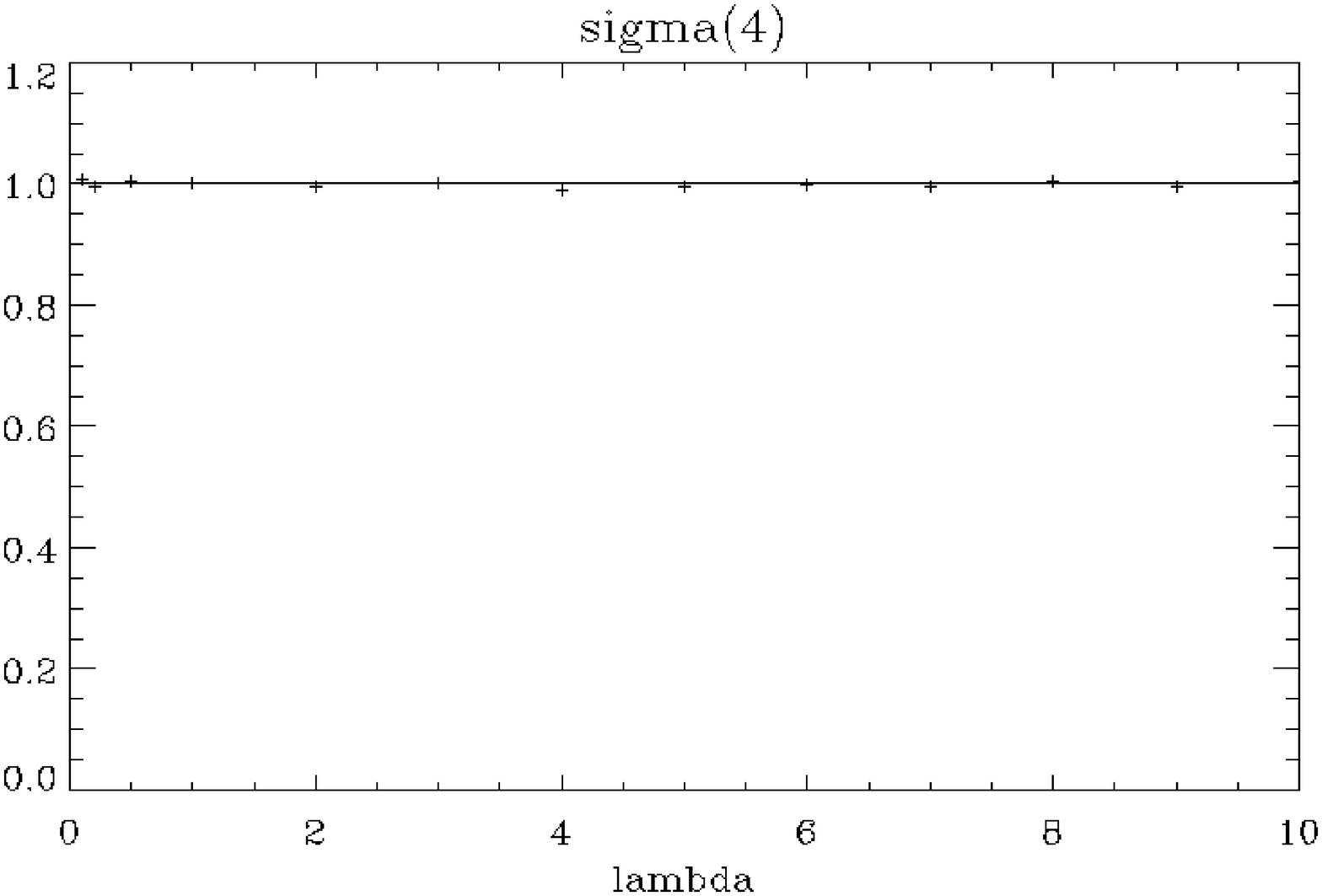}
\includegraphics[width=2.9in]{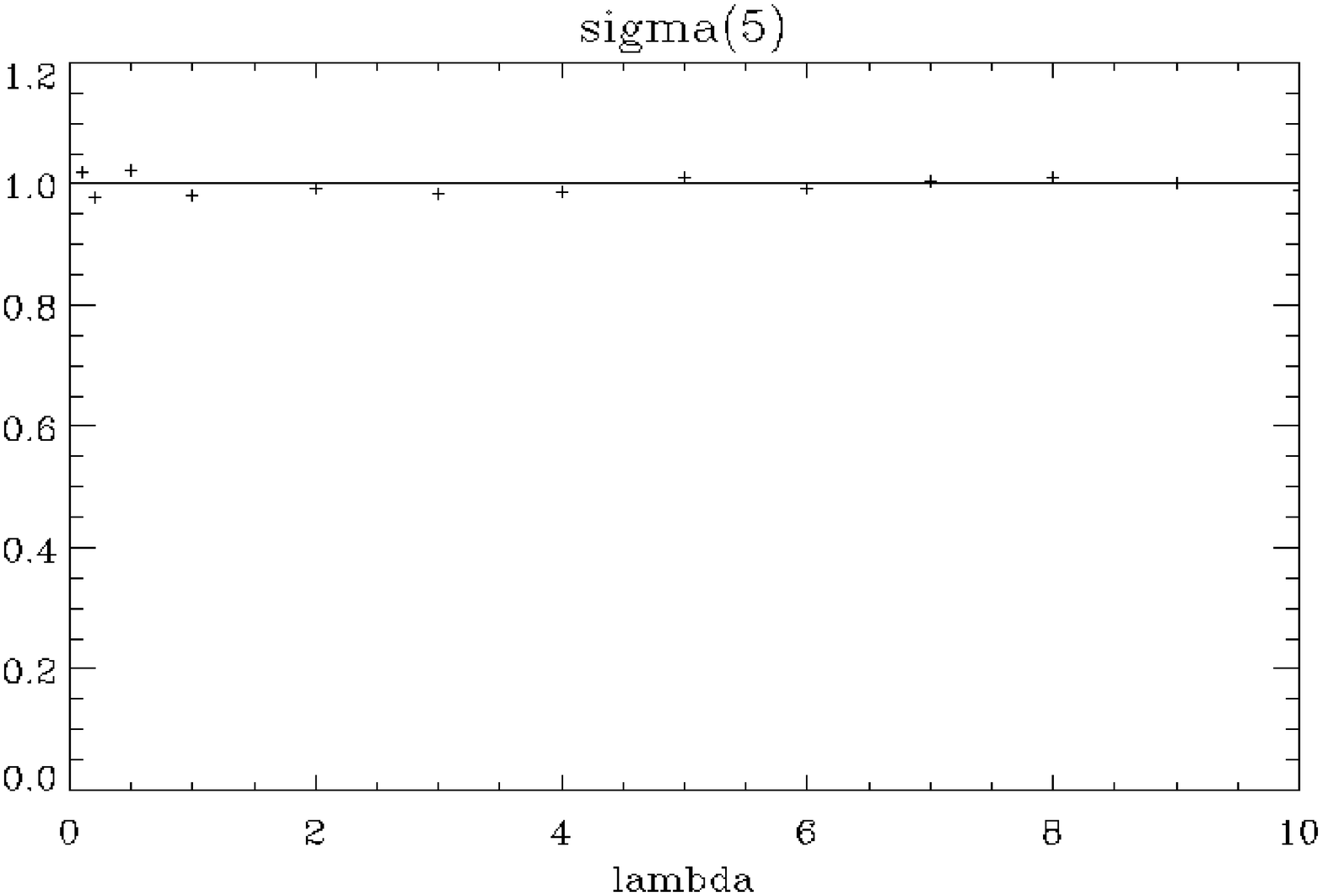}
\caption{Normalized value ($\mathbf{ (\sigma_{(j)})_{\text{simulated}} }/ (\sigma_{(j)})_{\text{theoretical}}$) of the stabilized variances at each scale $j$ as a function of $\lambda$.}
\label{sigma}
\end{figure*}

\begin{figure*}[h]
\centering
\includegraphics[width=2.9in]{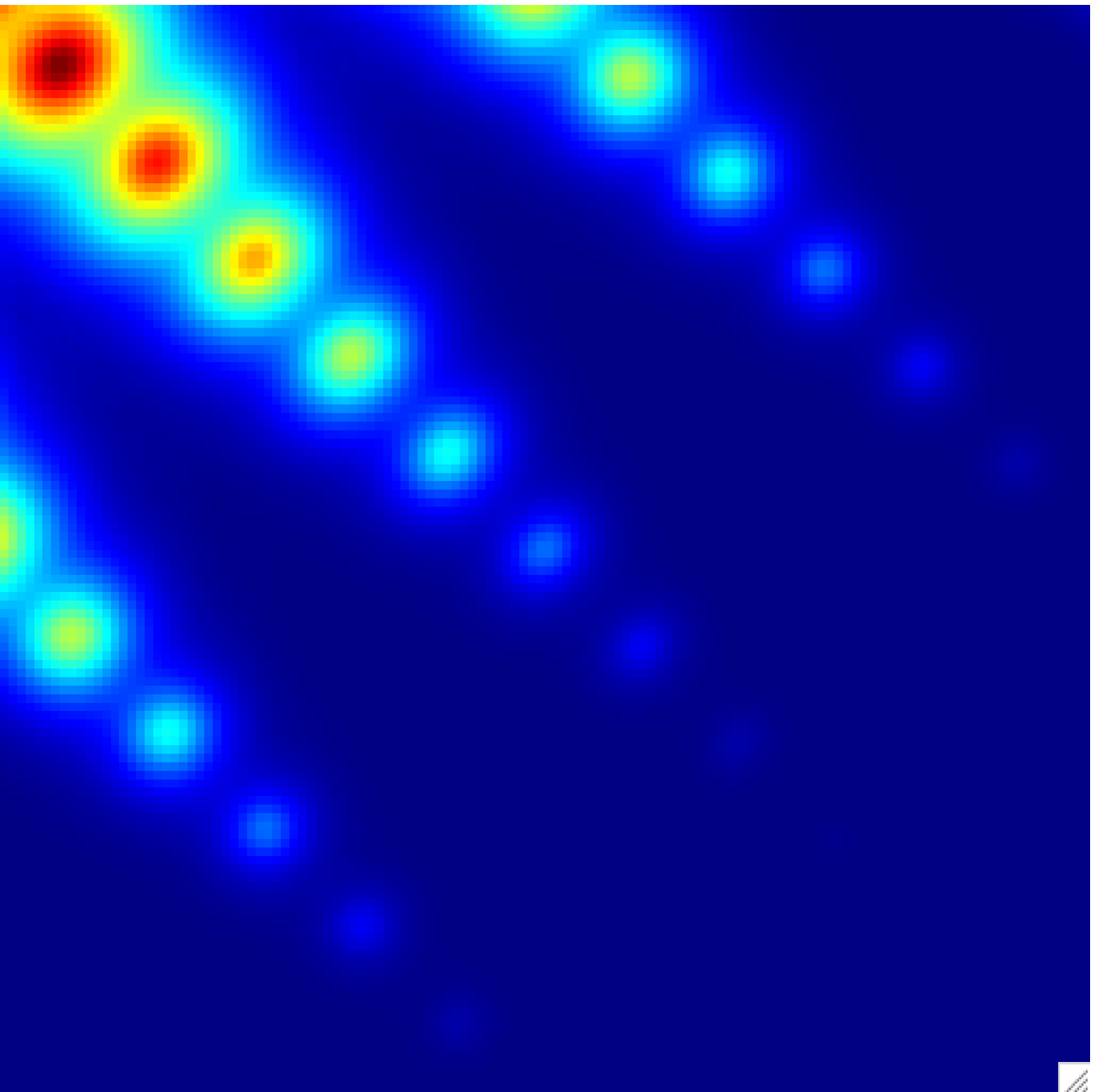}
\includegraphics[width=2.9in]{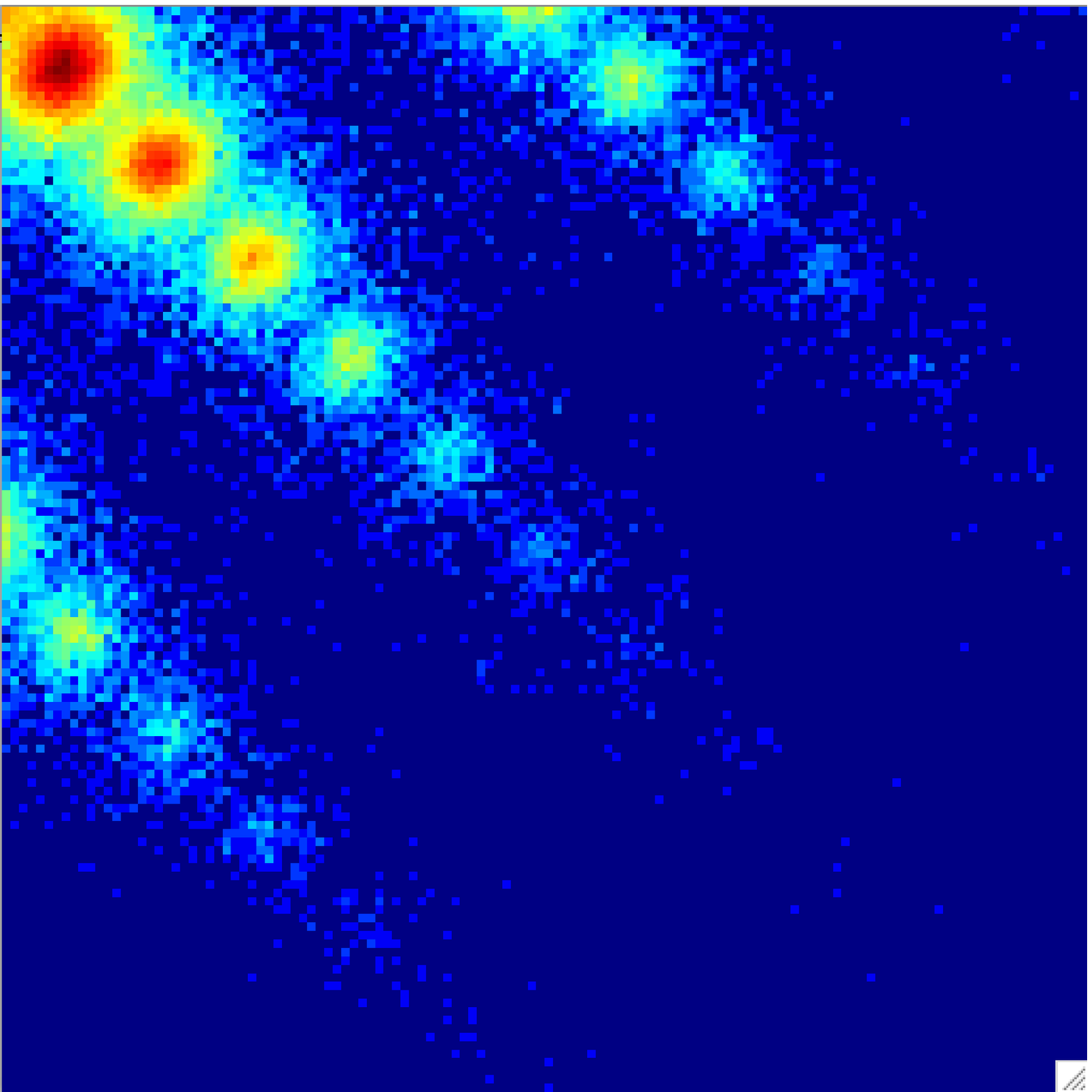}
\includegraphics[width=2.9in]{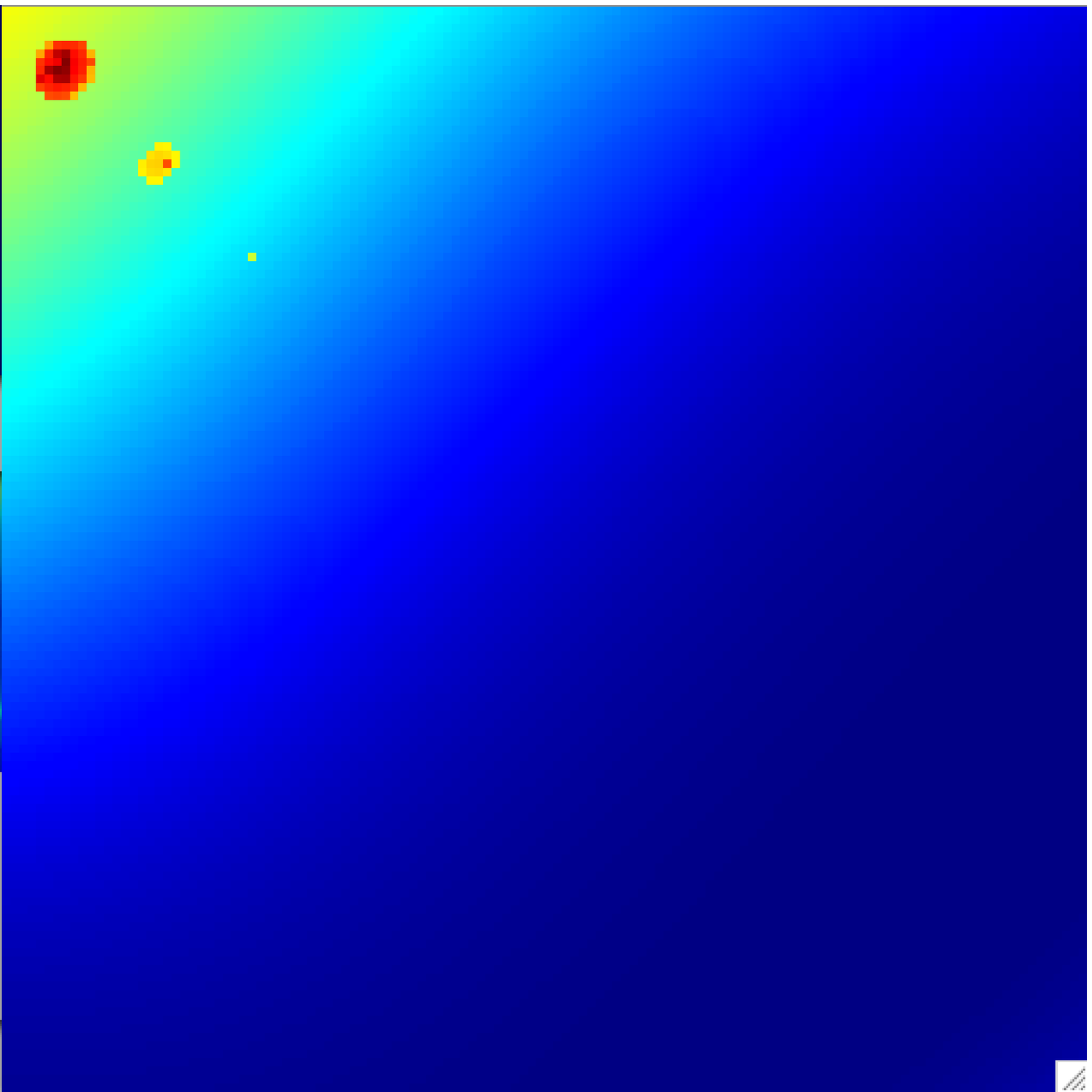}
\includegraphics[width=2.9in]{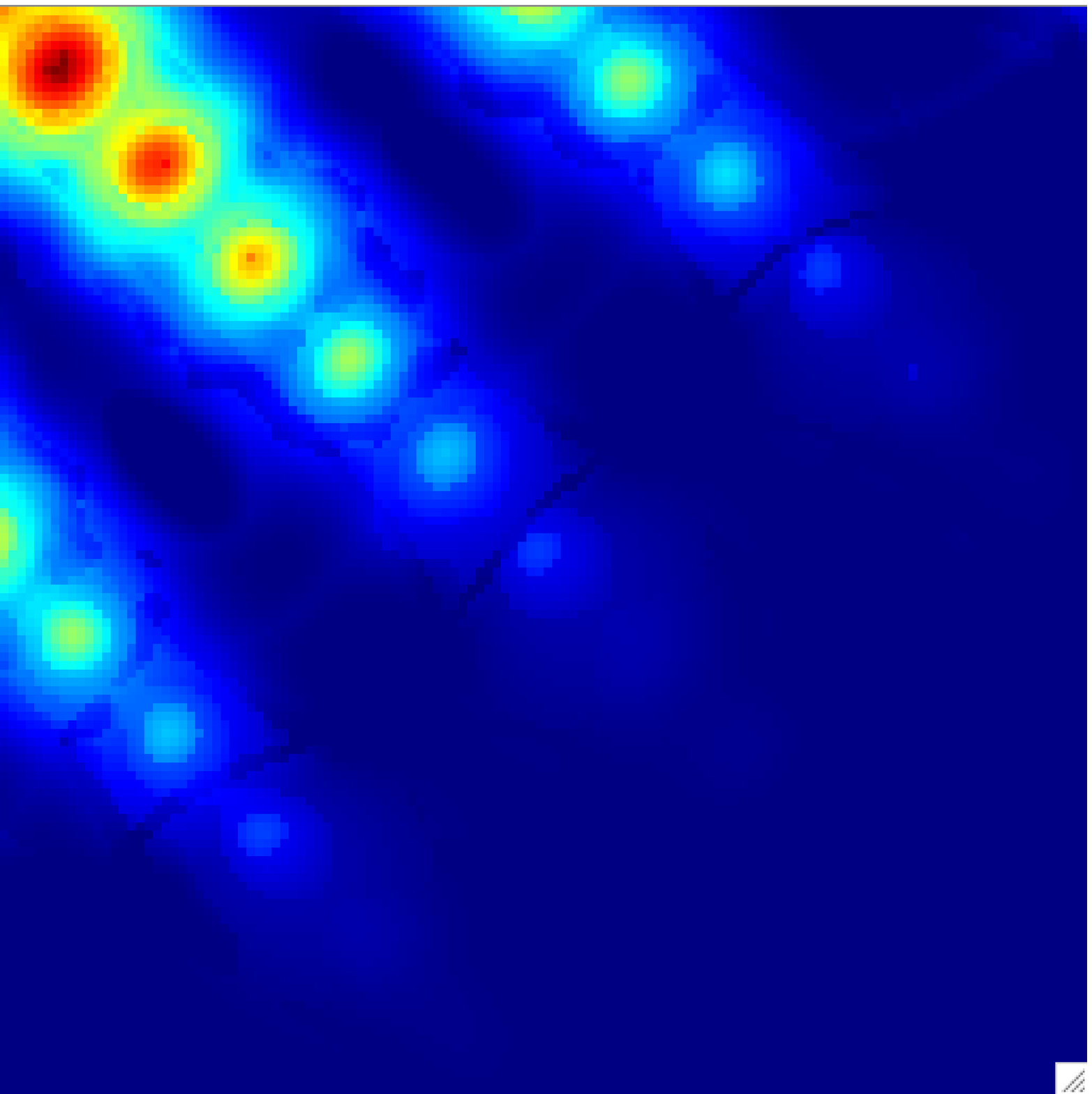}
\caption{Comparison of MS-VSTS with Anscombe + wavelet shrinkage on a single HEALPix face.
\emph{Top Left} : Sources of varying intensity.
\emph{Top Right} : Sources of varying intensity with Poisson noise.
\emph{Bottom Left} : Poisson sources of varying intensity reconstructed with Anscombe + wavelet shrinkage.
\emph{Bottom Right} : Poisson sources of varying intensity reconstructed with MS-VSTS.
}
\label{ansc}
\end{figure*}

We have simulated Poisson images of different constant intensities $\lambda$, computed the IUWT with MS-VSTS on each image and observed the variation of the normalized value of $\sigma_{(j)}$ ($\mathbf{ (\sigma_{(j)})_{\text{simulated}}} / (\sigma_{(j)})_{\text{theoretical}}$) as a function of $\lambda$ for each scale $j$ (Fig. \ref{sigma}). We see that the wavelet coefficients are stabilized when $\lambda \gtrsim 0.1$ except for the first wavelet scale, which is mostly constituted of noise. On Fig. \ref{ansc}, we compare the result of MS-VSTS with Anscombe + wavelet shrinkage, on sources of varying intensities. We see that MS-VSTS works well on sources of very low intensities, whereas Anscombe doesn't work when the intensity is too low.

\subsubsection{MS-VSTS + Curvelets}

As the first step of the algorithm is an IUWT, we can stabilize each resolution level as in Equation~(\ref{eq27}). We then apply the local ridgelet transform on each stabilized wavelet band.

It is not as straightforward as with the IUWT to derive the asymptotic noise variance in the stabilized curvelet domain. In our experiments, we derived them using simulated Poisson data of stationary intensity level $\lambda$. After having checked that the standard deviation in the curvelet bands becomes stabilized as the intensity level increases (which means that the stabilization is working properly), we stored the standard deviation $\sigma_{j,l}$ for each wavelet scale $j$ and each ridgelet band $l$ (Table~\ref{tabcurv}).

\begin{table*}[!h]
  \centering
   \caption{Asymptotic values of the variances $\sigma_{j,k}$ of the curvelet coefficients.
  }
  \begin{tabular}{|c|c|c|c|c|}
\hline
 $j$ & $l=1$ & $l=2$ & $l=3$ & $l=4$ \\
\hline
  1 & 1.74550 & 0.348175 & & \\
  2 & 0.230621 & 0.248233 & 0.196981 & \\
  3 & 0.0548140 & 0.0989918 & 0.219056 & \\
  4 & 0.0212912 & 0.0417454 & 0.0875663 & 0.20375 \\
  5 & 0.00989616 & 0.0158273 & 0.0352021 & 0.163248 \\
\hline
\end{tabular}
 
  \label{tabcurv}
\end{table*}


\section{Poisson Denoising}

\subsection{MS-VST + IUWT}

Under the hypothesis of homogeneous Poisson intensity, the stabilized wavelet coefficients $d_j$ behave like centered Gaussian variables of standard deviation $\sigma_{(j)}$. We can detect significant coefficients with binary hypothesis testing as in Gaussian denoising.

Under the null hypothesis $\mathcal{H}_0$ of homogeneous Poisson intensity, the distribution of the stabilized wavelet coefficient $d_j[k]$ at scale $j$ and location index $k$ can be written as:
\begin{equation}
\label{ }
p(d_j[k]) = \frac{1}{\sqrt{2\pi}\sigma_j}\exp(-d_j[k]^2 / 2 \sigma_j^2) .
\end{equation}

The rejection of the hypothesis $\mathcal{H}_0$ depends on the double-sided p-value:
\begin{equation}
\label{ }
p_j[k] = 2 \frac{1}{\sqrt{2\pi}\sigma_j}\int_{|d_j[k]|}^{+\infty} \exp(-x^2 / 2 \sigma_j^2) dx .
\end{equation}

Consequently, to accept or reject $\mathcal{H}_0$, we compare each $|d_j[k]|$ with a critical threshold $\kappa \sigma_j$, $\kappa= 3,4 \text{ or } 5$ corresponding respectively to significance levels. This amounts to deciding that:
\begin{itemize}
  \item if $|d_j[k]| \geqslant  \kappa \sigma_j$, $d_j[k]$ is significant.
  \item if $|d_j[k]| < \kappa \sigma_j$, $d_j[k]$ is not significant.
\end{itemize}

Then we have to invert the MS-VSTS scheme to reconstruct the estimate. However, although the direct inversion is possible (Eq. (\ref{eq30})), it can not guarantee a positive intensity estimate, while the Poisson intensity is always nonnegative. A positivity projection can be applied, but important structures could be lost in the estimate. To tackle this problem, we reformulate the reconstruction as a convex optimisation problem and solve it iteratively with an algorithm based on Hybrid Steepest Descent (HSD)~\citep{Yamada}.

We define the multiresolution support $\mathcal{M}$, which is determined by the set of detected significant coefficients after hypothesis testing:
\begin{equation}
\label{eq33}
\mathcal{M} := \{ (j,k) | \text{if } d_j[k] \text{ is declared significant} \} .
\end{equation}

We formulate the reconstruction problem as a convex constrained minimization problem:

\begin{equation}
\label{eq34}
\begin{split}
\text{Arg} \min_{\mathbf{X}} \| \mathbf{ \Phi}^{T}\mathbf{X}\|_1,
\text{s.t.} \\ \: \left\{\begin{array}{c}\mathbf{X} \geqslant 0 , \\\forall (j,k)\in \mathcal{M},      (\mathbf{ \Phi}^{T}\mathbf{X})_j[k]=(\mathbf{ \Phi}^{T}\mathbf{Y})_j[k] , \end{array}\right.
\end{split}
\end{equation}
where $\mathbf{\Phi}$ denotes the IUWT synthesis operator.

This problem is solved with the following iterative scheme: the image is initialised by $\mathbf{X}^{(0)} = 0$, and the iteration scheme is, for $n=0$ to $N_{\max}-1$:

\begin{eqnarray}
\tilde{\mathbf{X}} &=& P_{+}[\mathbf{ X}^{(n)} + \mathbf{ \Phi} P_{\mathcal{M}} \mathbf{ \Phi}^{T} (\mathbf{ Y} - \mathbf{ X}^{(n)})] \\
\mathbf{X}^{(n+1)} &=& \mathbf{ \Phi}\text{ST}_{\lambda_n}[\mathbf{ \Phi}^{T}\tilde{\mathbf{X}}]
\end{eqnarray}
where $P_{+}$ denotes the projection on the positive orthant, $P_{\mathcal{M}}$ denotes the projection on the multiresolution support $\mathcal{M}$:
\begin{equation}
P_{\mathcal{M}}d_j[k] = \left\{\begin{array}{cc} d_j[k] & \text{if} \  (j,k) \in \mathcal{M} , \\0 & \text{otherwise} \end{array} . \right.
\end{equation}
and $\text{ST}_{\lambda_n}$ the soft-thresholding with threshold $\lambda_n$:
\begin{equation}
\text{ST}_{\lambda_n} [d] = \left\{\begin{array}{cc} \mathrm{sign}(d)(|d| - \lambda_n) & \text{if} \ |d| \geqslant \lambda_n , \\0 & \text{otherwise} \end{array} . \right.
\end{equation}
We chose a decreasing threshold $\lambda_n = \frac{N_{\max} - n}{N_{\max} - 1},n=1,2,\cdots,N_{\max}$.

The final estimate of the Poisson intensity is: $\hat{\mathbf{\Lambda}} = \mathbf{X}^{(N_{\max})}$. Algorithm~\ref{alg1} summarizes the main steps of the MS-VSTS + IUWT denoising algorithm.

\begin{algorithm}[!h]
\caption{MS-VSTS + IUWT Denoising}
\label{alg1}
\begin{algorithmic}[1]
\REQUIRE $\quad$ data $a_0:=\mathbf{Y}$, number of iterations $N_{\max}$, threshold $\kappa$ \\
\underline{\emph{\textbf{Detection}}} \\
\FOR{$j=1$ to $J$}
\STATE Compute $a_j$ and $d_j$ using (\ref{eq27}).
\STATE Hard threshold $|d_j[k]|$ with threshold $\kappa \sigma_j$ and update $\mathcal{M}$.
\ENDFOR \\
\underline{\emph{\textbf{Estimation}}} \\
\STATE Initialize $\mathbf{X}^{(0)}=0$, $\lambda_0 = 1$.
\FOR{$n=0$ to $N_{\max}-1$}
\STATE $\tilde{\mathbf{X}}= P_{+}[\mathbf{ X}^{(n)} + \mathbf{ \Phi} P_{\mathcal{M}} \mathbf{ \Phi}^{T} (\mathbf{ Y} - \mathbf{ X}^{(n)})]$.
\STATE $\mathbf{X}^{(n+1)} = \mathbf{ \Phi}\text{ST}_{\lambda_n}[\mathbf{ \Phi}^{T}\tilde{\mathbf{X}}]$.
\STATE $\lambda_{n+1} = \frac{N_{\max} - (n+1)}{N_{\max} - 1}$.
\ENDFOR
\STATE Get the estimate $\hat{\mathbf{\Lambda}} = \mathbf{X}^{(N_{\max})}$.

\end{algorithmic}
\end{algorithm}

\subsection{Multi-resolution support adaptation}

When two sources are too close, the \textbf{less} intense source may not be detected because of the negative wavelet coefficients of the brightest source. To avoid such a drawback, we may update the multi-resolution support at each iteration. The idea is to withdraw the detected sources and to make a detection on the remaining residual, so as to detect the sources which may have been missed at the first detection.

At each iteration $n$, we compute the MS-VSTS of $\mathbf{X}^{(n)}$. We denote $d^{(n)}_j[k]$ the stabilised coefficients of $\mathbf{X}^{(n)}$. We make a hard thresholding on $(d_j[k]-d^{(n)}_j[k])$ with the same thresholds as in the detection step. Significant coefficients are added to the multiresolution support $\mathcal{M}$.

\begin{algorithm}
\caption{MS-VSTS + IUWT Denoising + Multiresolution Support Adaptation}
\label{alg4}
\begin{algorithmic}[1]
\REQUIRE $\quad$ data $a_0:=\mathbf{Y}$, number of iterations $N_{\max}$, threshold $\kappa$ \\
\underline{\emph{\textbf{Detection}}} \\
\FOR{$j=1$ to $J$}
\STATE Compute $a_j$ and $d_j$ using (\ref{eq27}).
\STATE Hard threshold $|d_j[k]|$ with threshold $\kappa \sigma_j$ and update $\mathcal{M}$.
\ENDFOR \\
\underline{\emph{\textbf{Estimation}}} \\
\STATE Initialize $\mathbf{X}^{(0)}=0$, $\lambda_0 = 1$.
\FOR{$n=0$ to $N_{\max}-1$}
\STATE $\tilde{\mathbf{X}}= P_{+}[\mathbf{ X}^{(n)} + \mathbf{ \Phi} P_{\mathcal{M}} \mathbf{ \Phi}^{T} (\mathbf{ Y} - \mathbf{ X}^{(n)})]$.
\STATE $\mathbf{X}^{(n+1)} = \mathbf{ \Phi}\text{ST}_{\lambda_n}[\mathbf{ \Phi}^{T}\tilde{\mathbf{X}}]$.
\STATE Compute the MS-VSTS on  $\mathbf{X}^{(n)}$ to get the stabilised coeffcients $d^{(n)}_j$.
\STATE Hard threshold $|d_j[k]-d^{(n)}_j[k]|$ and update $\mathcal{M}$.
\STATE $\lambda_{n+1} = \frac{N_{\max} - (n+1)}{N_{\max} - 1}$.
\ENDFOR
\STATE Get the estimate $\hat{\mathbf{\Lambda}} = \mathbf{X}^{(N_{\max})}$.

\end{algorithmic}
\end{algorithm}

The main steps of the algorithm are summarized in Algorithm~\ref{alg4}. In practice, we use Algorithm~\ref{alg4} instead of Algorithm~\ref{alg1} in our experiments.

\subsection{MS-VST + Curvelets}

Insignificant coefficients are zeroed by using the same hypothesis testing framework as in the wavelet scale. At each wavelet scale $j$ and ridgelet band $k$, we make a hard thresholding on curvelet coefficients with threshold $\kappa \sigma_{j,k}$, $\kappa= 3,4 \text{ or } 5$. Finally, a direct reconstruction can be performed by first inverting the local ridgelet transforms and then inverting the MS-VST + IUWT~(Equation~(\ref{eq30})). An iterative reconstruction may also be performed.

Algorithm~\ref{algcurv} summarizes the  main steps of the MS-VSTS + Curvelets denoising algorithm.

\begin{algorithm}
\caption{MS-VSTS + Curvelets Denoising}
\label{algcurv}
\begin{algorithmic}[1]
\STATE Apply the MS-VST + IUWT with $J$ scales to get the stabilized wavelet subbands $d_j$.
\STATE Set $B_1 = B_{\min}$.
\FOR{$j=1$ to $J$}
\STATE Partition the subband $d_j$ with blocks of side-length $B_j$ and apply the digital ridgelet transform to each block to obtain the stabilized curvelets coefficients.
\IF {$j$ modulo $2=1$}
\STATE $B_{j+1} = 2 B_j$
\ELSE
\STATE $B_{j+1} =  B_j$
\ENDIF \\
\STATE HTs on the stabilized curvelet coefficients.
\ENDFOR \\
\STATE Invert the ridgelet transform in each block before inverting the MS-VST + IUWT.

\end{algorithmic}
\end{algorithm}

\subsection{Experiments}

The method was tested on simulated Fermi data. The simulated data are the sum of a Milky Way diffuse background model and 1000 gamma ray point sources. We based our Galactic diffuse emission model intensity on the model $gll\_iem\_v02$ obtained at the Fermi Science Support Center~\citep{Models}
. This model results from a fit of the LAT photons with various gas templates as well as inverse Compton in several energy bands. We used a realistic point-spread function for the sources, based on Monte Carlo simulations of the LAT and accelerator tests, that scale approximately as $0.8(E/1GeV)^{-0.8}$ degrees. The position of the 205 brightest sources were taken from the Fermi 3-month source list~\citep{Abdo}. The position of the 795 remaining sources follow the LAT 1-year Point Source Catalog~\citep{Catalog}
  sources distribution: each simulated source was randomly sorted in a box of $\Delta$l=5$^o$ and $\Delta$b=1$^o$ around a LAT 1-year catalog source. We simulated each source assuming a power-law dependence with its spectral index given by the 3-month source list and the first year catalog. We used an exposure of $3.10^{10} s.cm^2$ corresponding approximatively to one year of Fermi all-sky survey around 1 GeV. The simulated counts map shown here correspond to photons energy from 150 MeV to 20 GeV.

\begin{figure*}
\begin{center}
\includegraphics[width=2.9in]{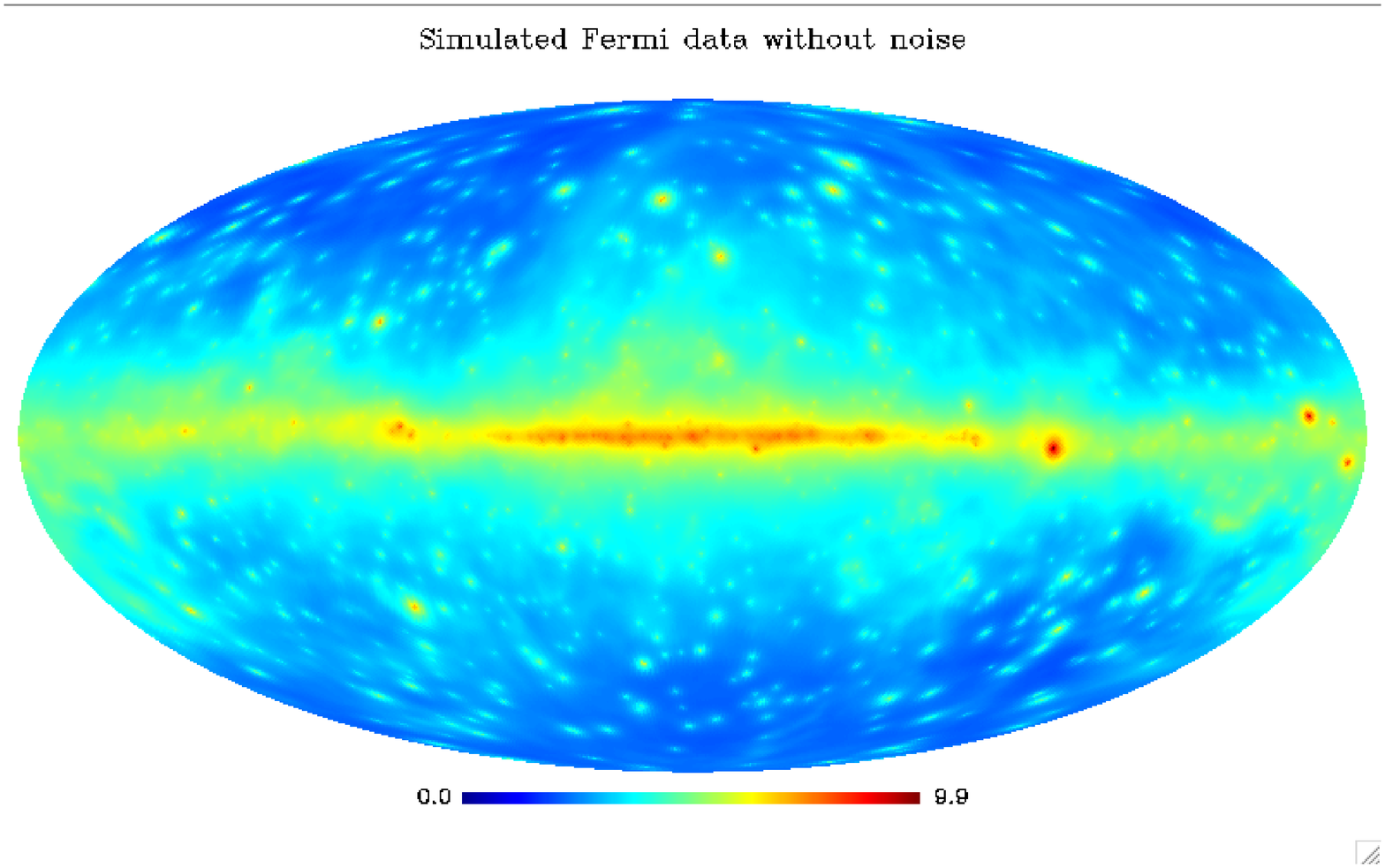} \hfill
\includegraphics[width=2.9in]{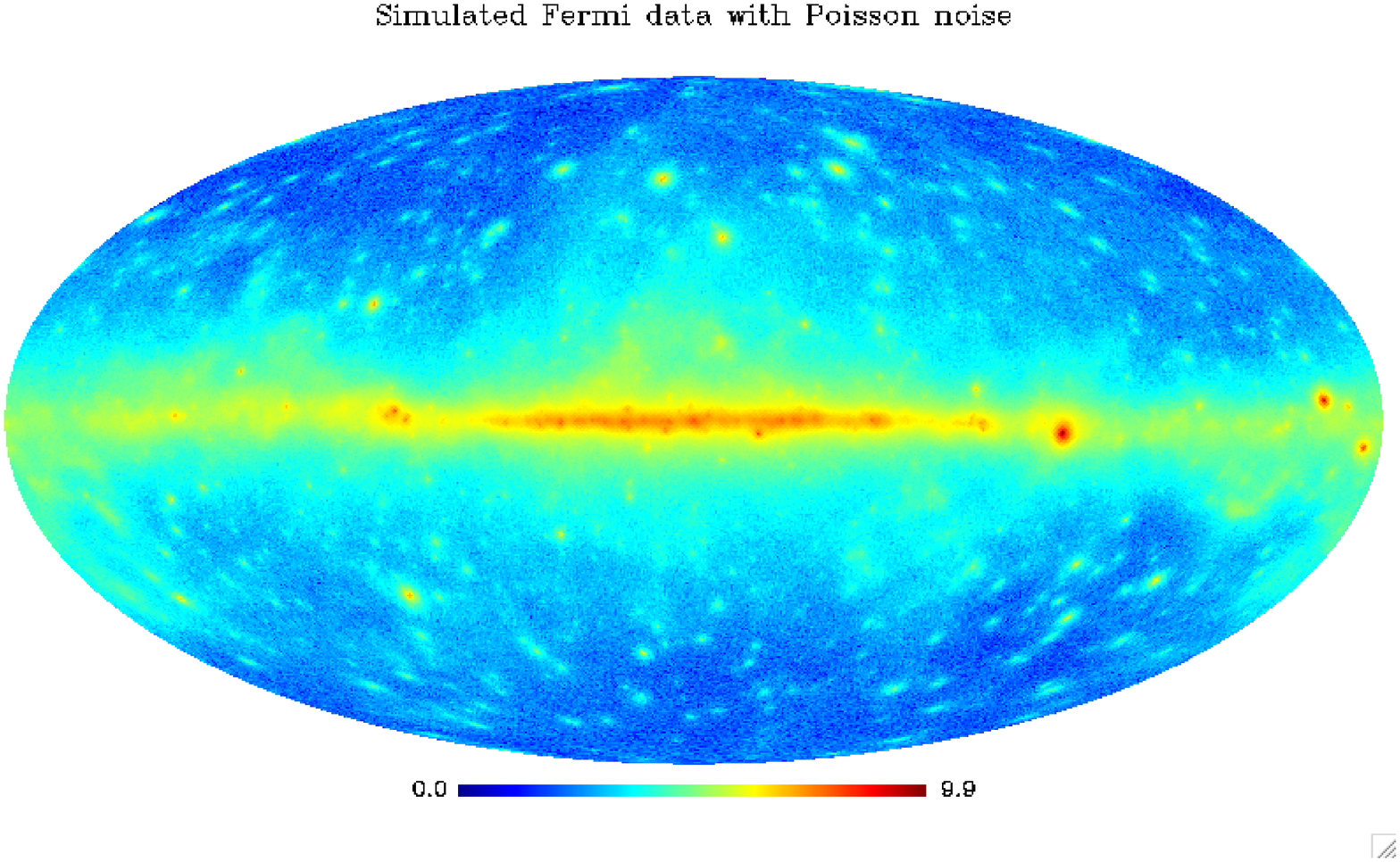} \hfill
\includegraphics[width=2.9in]{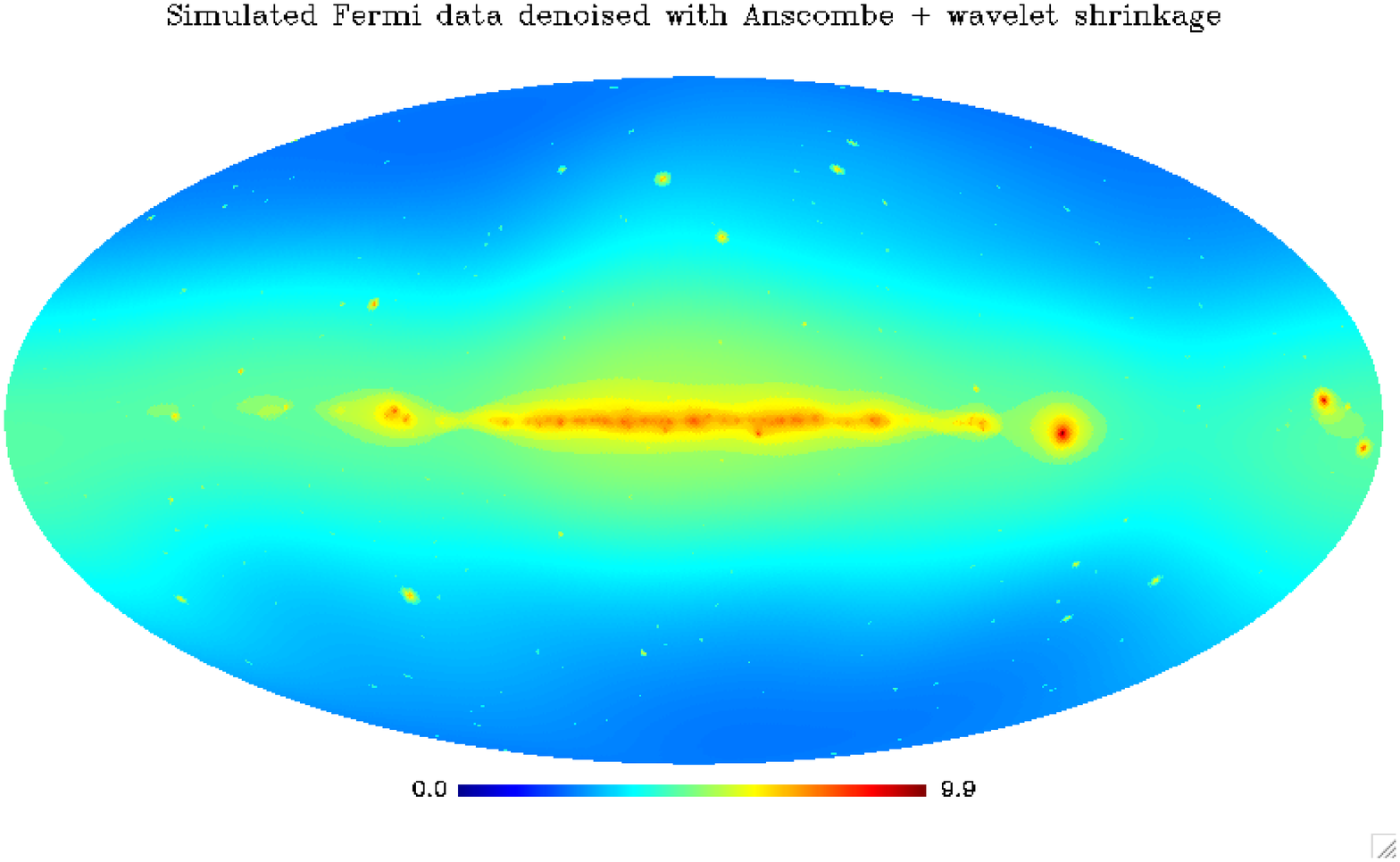} \hfill
\includegraphics[width=2.9in]{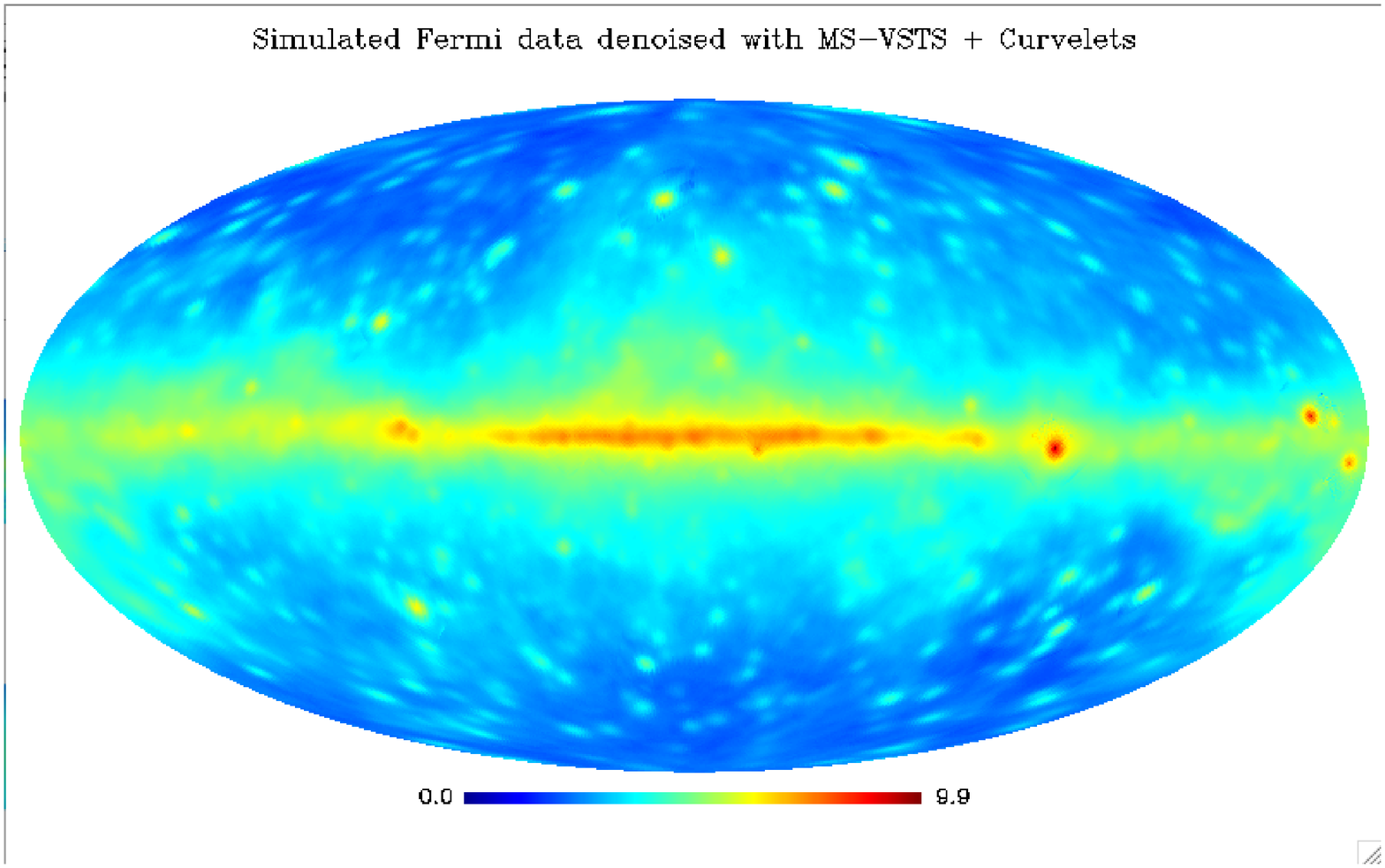} \hfill
\includegraphics[width=2.9in]{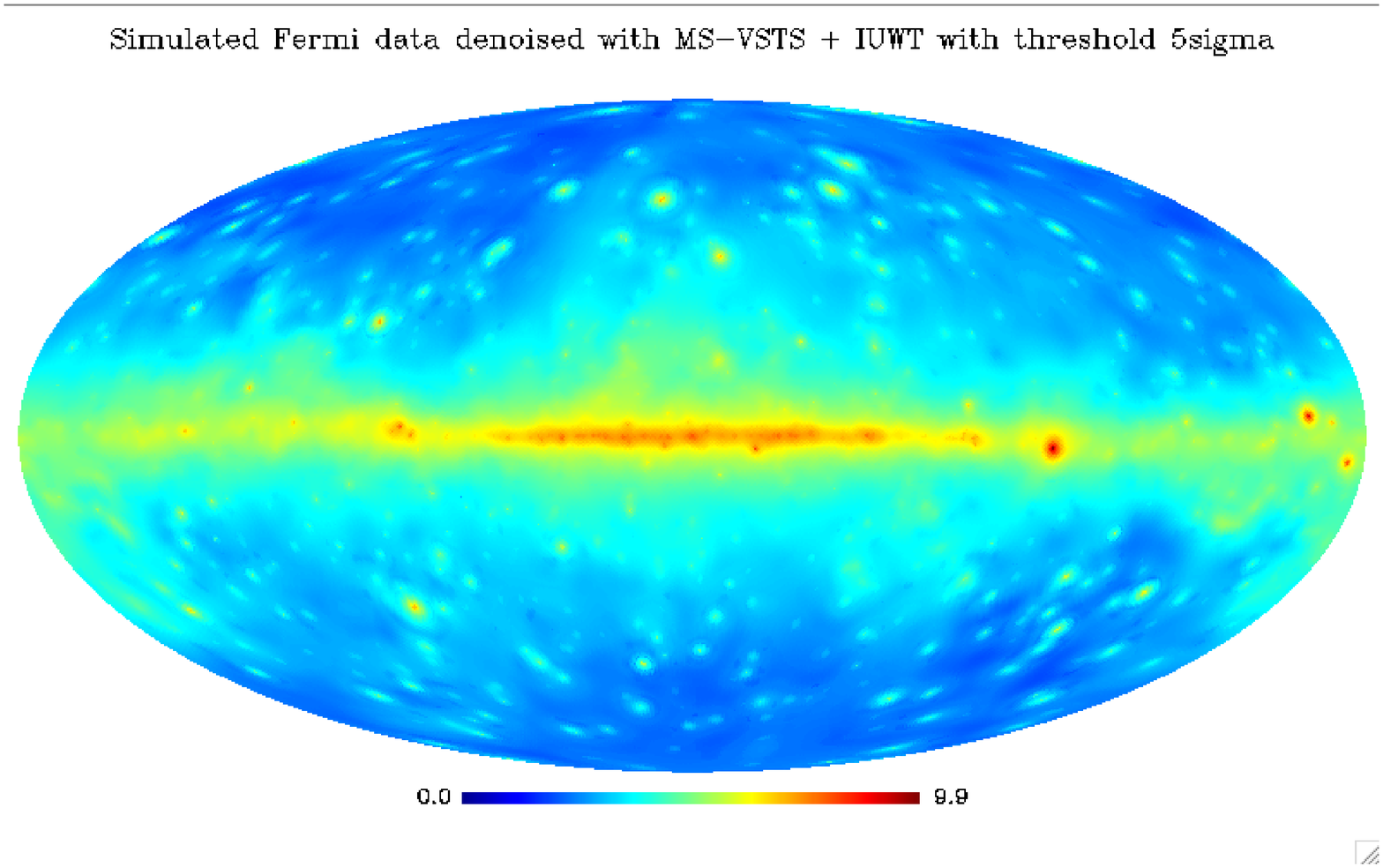} \hfill
\includegraphics[width=2.9in]{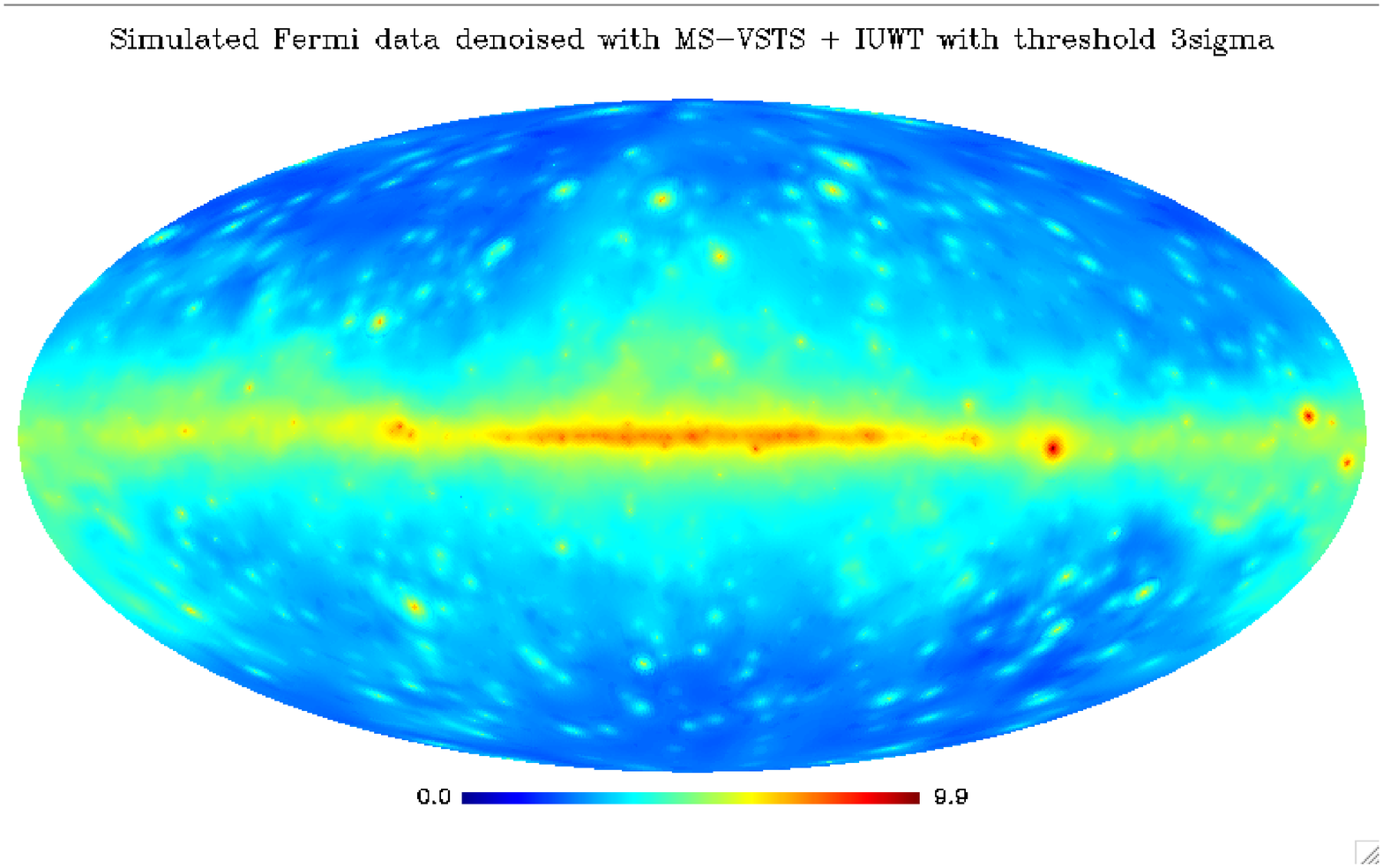}

\caption{\emph{Top Left}: Fermi simulated map without noise.
\emph{Top Right}: Fermi simulated map with Poisson noise.
\emph{Middle Left}: Fermi simulated map denoised with Anscombe VST + wavelet shrinkage.
\emph{Middle Right}: Fermi simulated map denoised with MS-VSTS + curvelets (Algorithm~\ref{algcurv}).
\emph{Bottom Left}: Fermi simulated map denoised with MS-VSTS + IUWT (Algorithm~\ref{alg1}) with threshold $5\sigma_j$.
\emph{Bottom Right}: Fermi simulated map denoised with MS-VSTS + IUWT (Algorithm~\ref{alg1}) with threshold $3\sigma_j$.
Pictures are in logarithmic scale.}
\label{rechsd}
\end{center}
\end{figure*}

\begin{figure*}
\begin{center}
\includegraphics[width=2.9in]{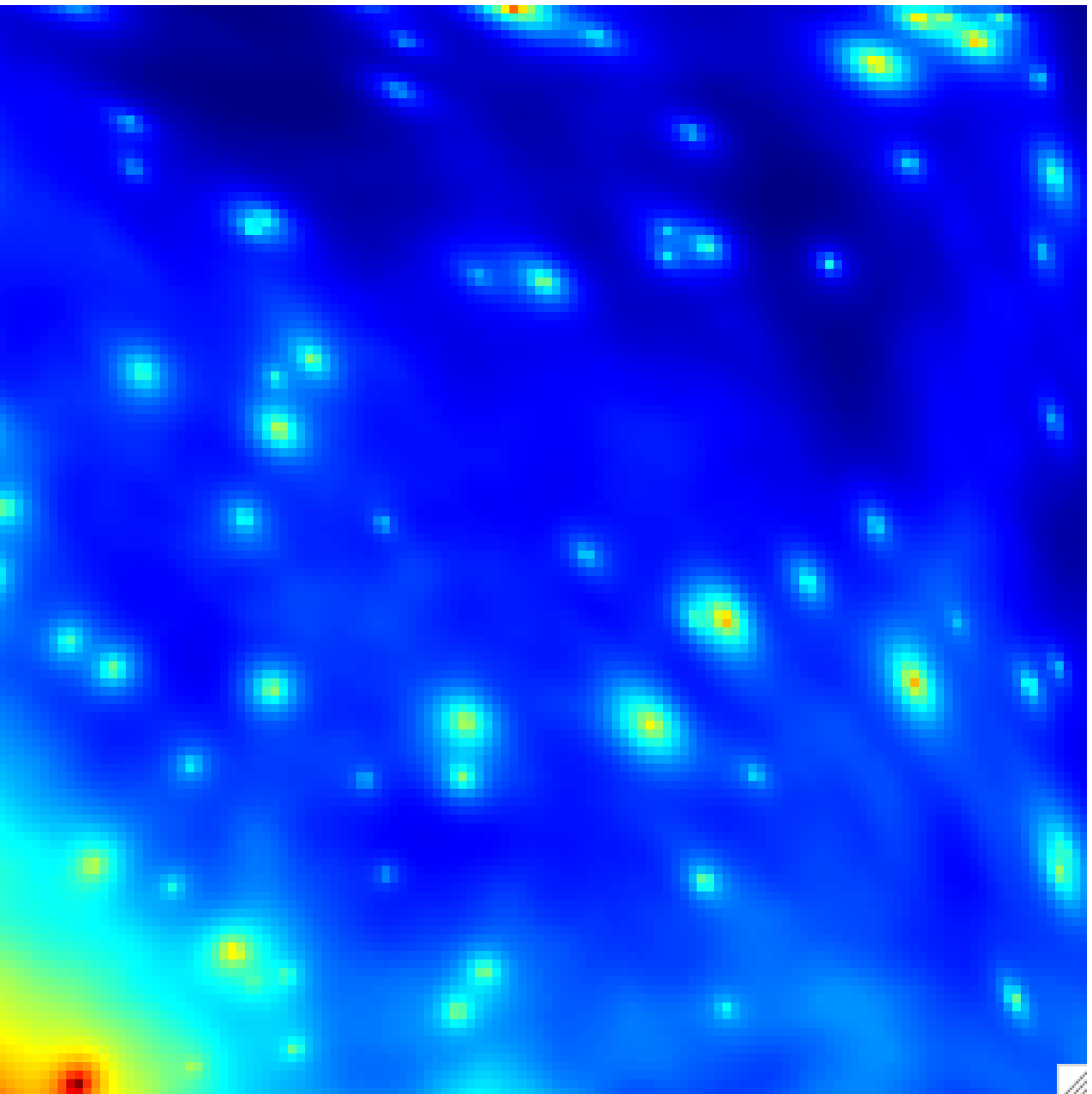} \hfill
\includegraphics[width=2.9in]{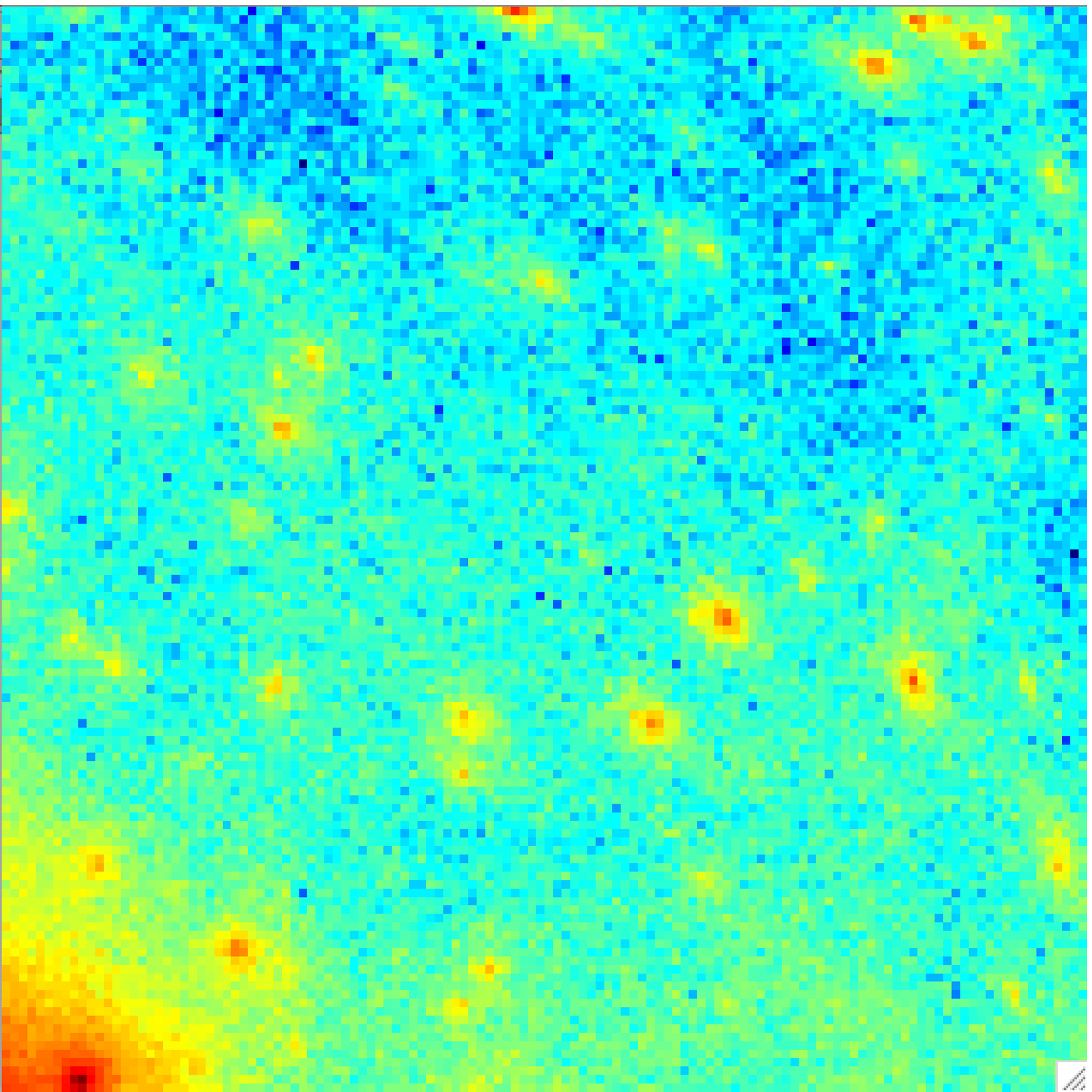} \hfill
\includegraphics[width=2.9in]{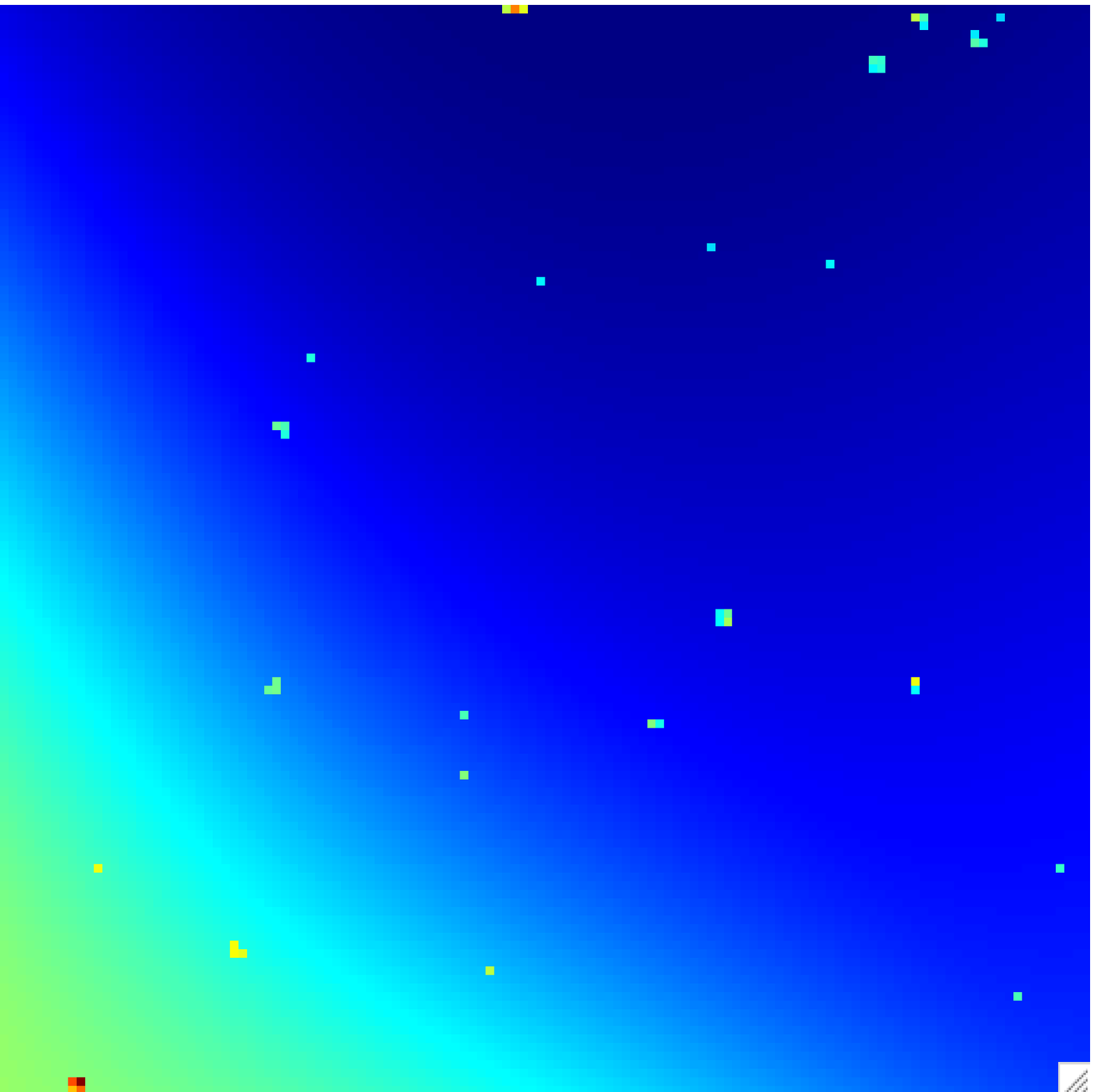} \hfill
\includegraphics[width=2.9in]{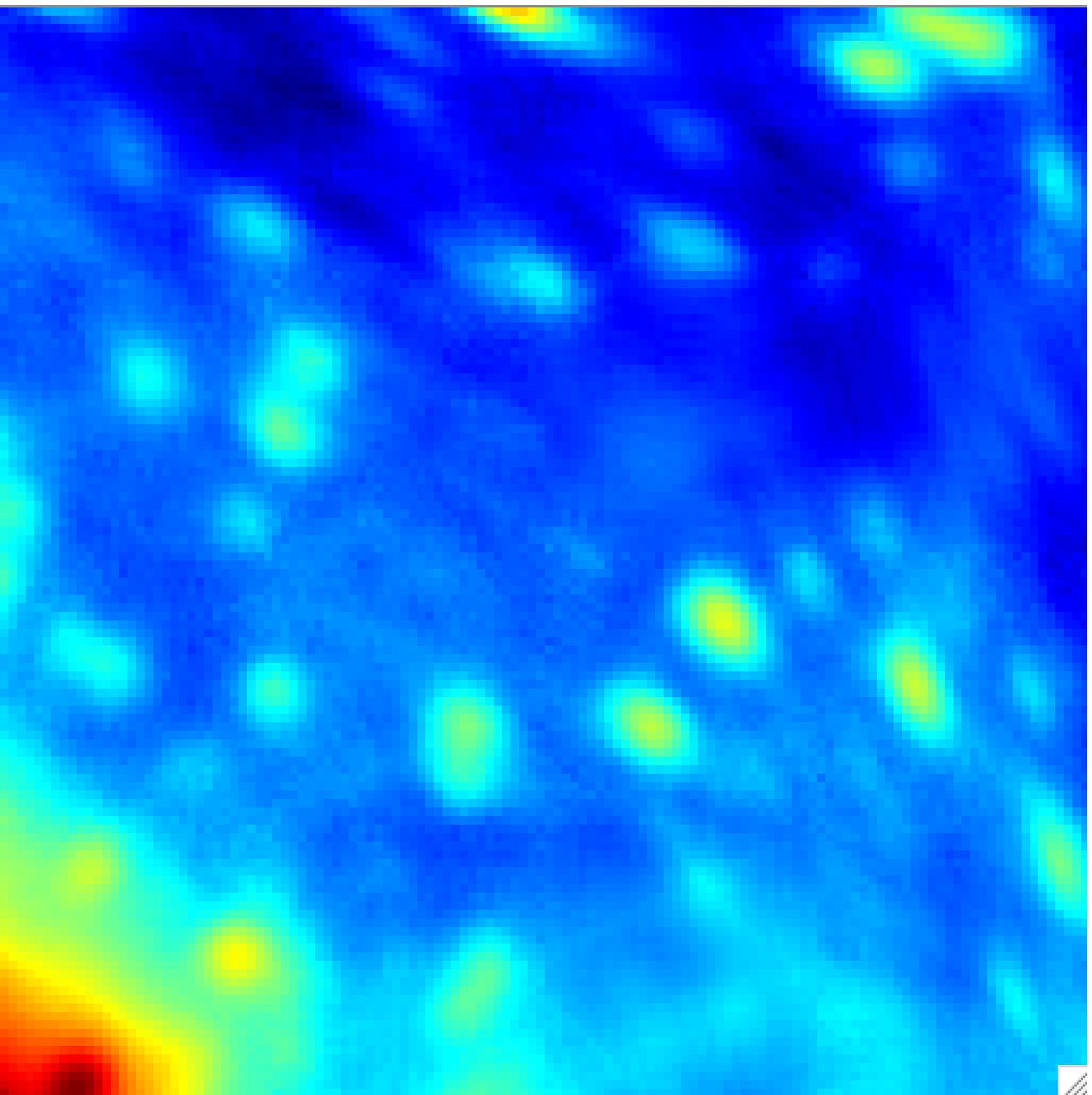} \hfill
\includegraphics[width=2.9in]{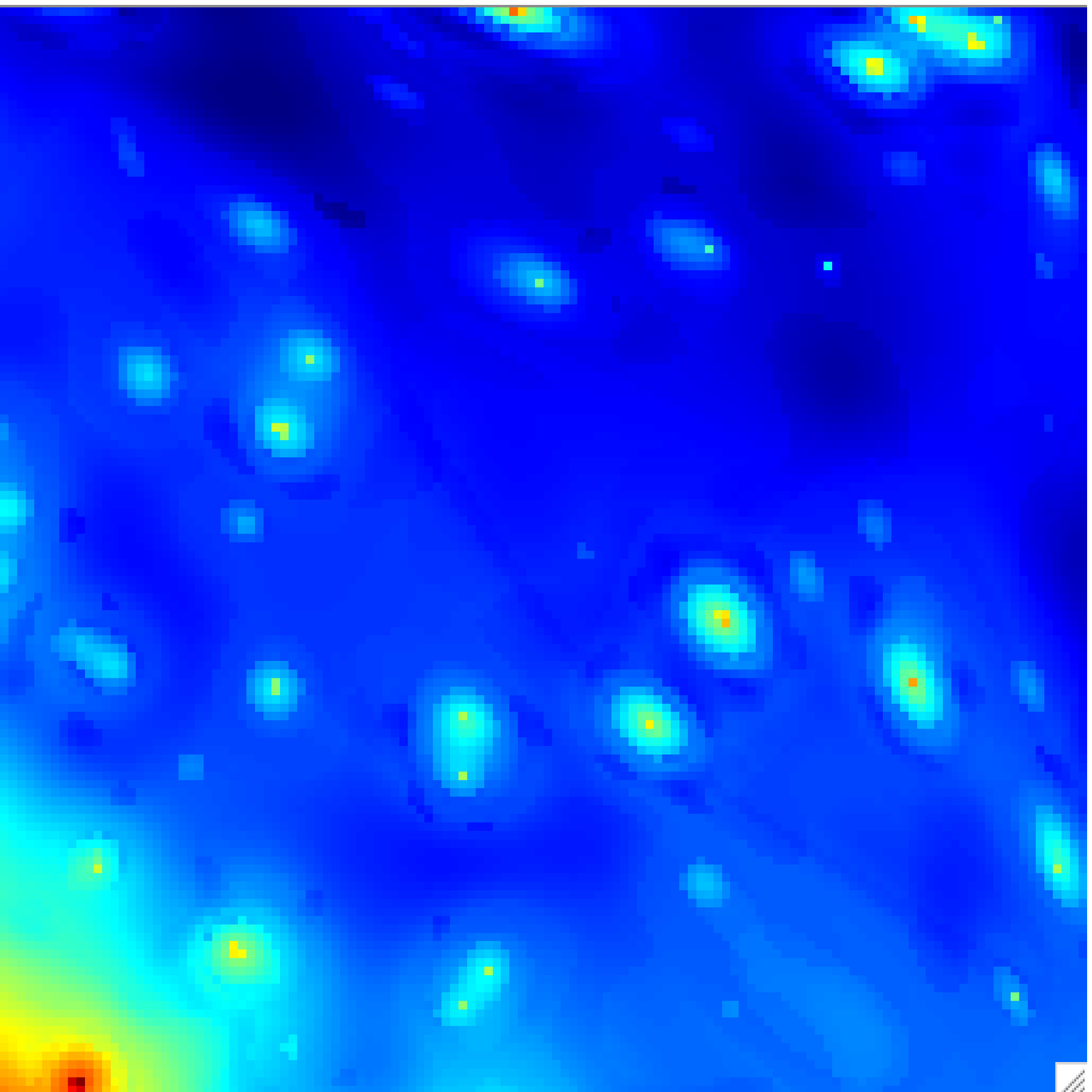} \hfill
\includegraphics[width=2.9in]{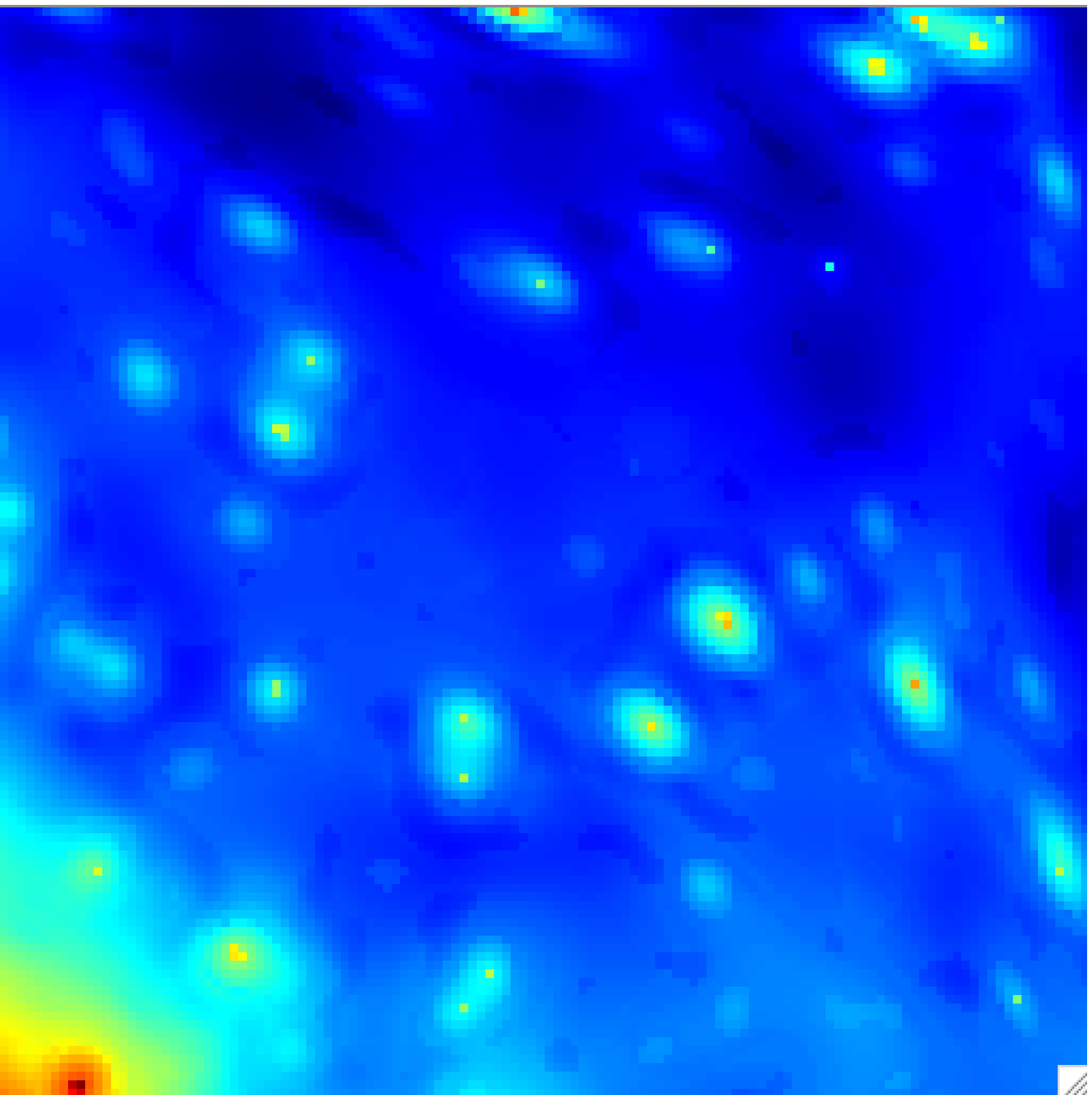} \hfill
\caption{View of a single HEALPix face from the results of Figure~\ref{rechsd}.
\emph{Top Left}: Fermi simulated map without noise.
\emph{Top Right}: Fermi simulated map with Poisson noise.
\emph{Middle Left}: Fermi simulated map denoised with Anscombe VST + wavelet shrinkage.
\emph{Middle Right}: Fermi simulated map denoised with MS-VSTS + curvelets (Algorithm~\ref{algcurv}).
\emph{Bottom Left}: Fermi simulated map denoised with MS-VSTS + IUWT (Algorithm~\ref{alg1}) with threshold $5\sigma_j$.
\emph{Bottom Right}: Fermi simulated map denoised with MS-VSTS + IUWT (Algorithm~\ref{alg1}) with threshold $3\sigma_j$.
Pictures are in logarithmic scale.
}
\label{recface}
\end{center}
\end{figure*}

Fig.~\ref{rechsd} compares the result of denoising with MS-VST + IUWT (Algorithm~\ref{alg1}), MS-VST + curvelets (Algorithm~\ref{algcurv}) and Anscombe VST + wavelet shrinkage on a simulated Fermi map. Fig.~\ref{recface} shows one HEALPix face of the results. 
As expected from theory, the Anscombe method produces poor results to  denoise Fermi data, because the underlyning intensity is too weak. 
Both wavelet and curvelet denoising on the sphere  perform much better. 
For this application, wavelets are slightly better than curvelets ($SNR_{wavelets} = 65.8 dB$, $SNR_{curvelets} = 37.3 dB$, $SNR (dB) = 20 \log (\sigma_{signal} / \sigma_{noise})$). As this image contains many point sources, thisresult is expected. Indeed wavelet are better than curvelets to represent isotropic objects.

\section{Milky Way diffuse background study: denoising and inpainting}

In order to extract from the Fermi photon maps the galactic diffuse emission, we want to remove the  point sources from the Fermi image. As our HSD algorithm is very close to the MCA algorithm~\citep{starck2004}, an idea is to mask the most intense sources and to modify our algorithm in order to interpolate through the gaps exactly as in the MCA-Inpainting algorithm~\citep{Abrial}. This modified algorithm can be called MS-VSTS-Inpainting algorithm.

The problem can be reformulated as a convex constrained minimization problem:

\begin{equation}
\label{eq34}
\begin{split}
\text{Arg} \min_{\mathbf{X}} \| \mathbf{ \Phi}^{T}\mathbf{X}\|_1,
\text{s.t.} \\ \: \left\{\begin{array}{c}\mathbf{X} \geqslant 0 , \\\forall (j,k)\in \mathcal{M},      (\mathbf{ \Phi}^{T}\Pi \mathbf{X})_j[k]=(\mathbf{ \Phi}^{T} \mathbf{Y})_j[k] , \end{array}\right. 
\end{split}
\end{equation}
where $\Pi$ is a binary mask ($1$ on valid data and $0$ on invalid data).

The iterative scheme can be adapted to cope with a binary mask, which gives:
\begin{eqnarray}
\tilde{\mathbf{X}} = P_{+}[\mathbf{ X}^{(n)} + \mathbf{ \Phi} P_{\mathcal{M}} \mathbf{ \Phi}^{T} \Pi (\mathbf{ Y} - \mathbf{ X}^{(n)})] , \\
\mathbf{X}^{(n+1)} = \mathbf{ \Phi} \text{ST}_{\lambda_n}[\mathbf{ \Phi}\tilde{\mathbf{X}}] .
\end{eqnarray}

The thresholding strategy has to be adapted. Indeed, for the impainting task we need to have a very large initial threshold in order to have a very smooth image in the beginning and to refine the details progressively. We chose an exponentially decreasing threshold:
\begin{equation}
\label{eq42}
\lambda_{n} = \lambda_{\max}  (2^{(\frac{N_{\max} - n}{N_{\max} - 1})} -1),n=1,2,\cdots,N_{\max} ,
\end{equation}
where $\lambda_{\max} = \max (\mathbf{\Phi}^{T}\mathbf{X})$.

\begin{algorithm}
\caption{MS-VST + IUWT Denoising + Inpainting}
\label{alg2}

\begin{algorithmic}[1]
\REQUIRE $\quad$ data $a_0:=\mathbf{Y}$, mask $\Pi$, number of iterations $N_{\max}$, threshold $\kappa$.\\
\underline{\emph{\textbf{Detection}}} \\
\FOR{$j=1$ to $J$}
\STATE Compute $a_j$ and $d_j$ using (\ref{eq27}).
\STATE Hard threshold $|d_j[k]|$ with threshold $\kappa \sigma_j$ and update $\mathcal{M}$.
\ENDFOR \\
\underline{\emph{\textbf{Estimation}}} \\
\STATE Initialize $\mathbf{X}^{(0)}=0$, $\lambda_{0} = \lambda_{\max}$.
\FOR{$n=0$ to $N_{\max}-1$}
\STATE $\tilde{\mathbf{X}}= P_{+}[\mathbf{ X}^{(n)} + \mathbf{ \Phi} P_{\mathcal{M}} \mathbf{ \Phi}^{T} \Pi(\mathbf{ Y} - \mathbf{ X}^{(n)})]$.
\STATE $\mathbf{X}^{(n+1)} = \mathbf{ \Phi}^\text{ST}_{\lambda_n}[\mathbf{ \Phi}^{T}\tilde{\mathbf{X}}]$.
\STATE $\lambda_{n+1} = \lambda_{\max}  (2^{(\frac{N_{\max} - (n+1)}{N_{\max} - 1})} -1)$
\ENDFOR
\STATE Get the estimate $\hat{\mathbf{\Lambda}} = \mathbf{X}^{(N_{\max})}$.

\end{algorithmic}
\end{algorithm}

\subsection*{Experiment}

We applied this method on simulated Fermi data where we masked the most luminous sources.

\begin{figure*}
\begin{center}
\includegraphics[width=2.9in]{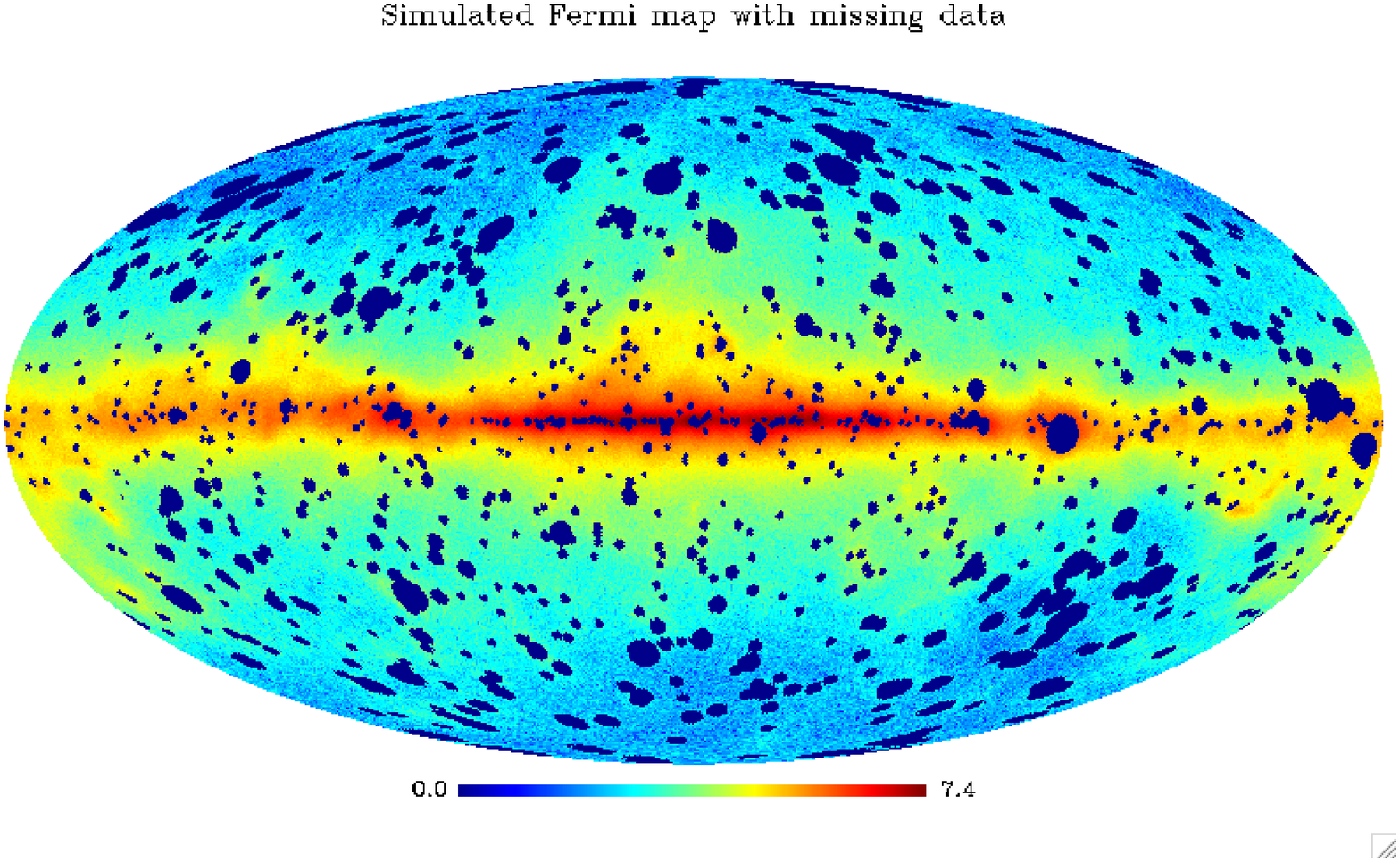} \hfill
\includegraphics[width=2.9in]{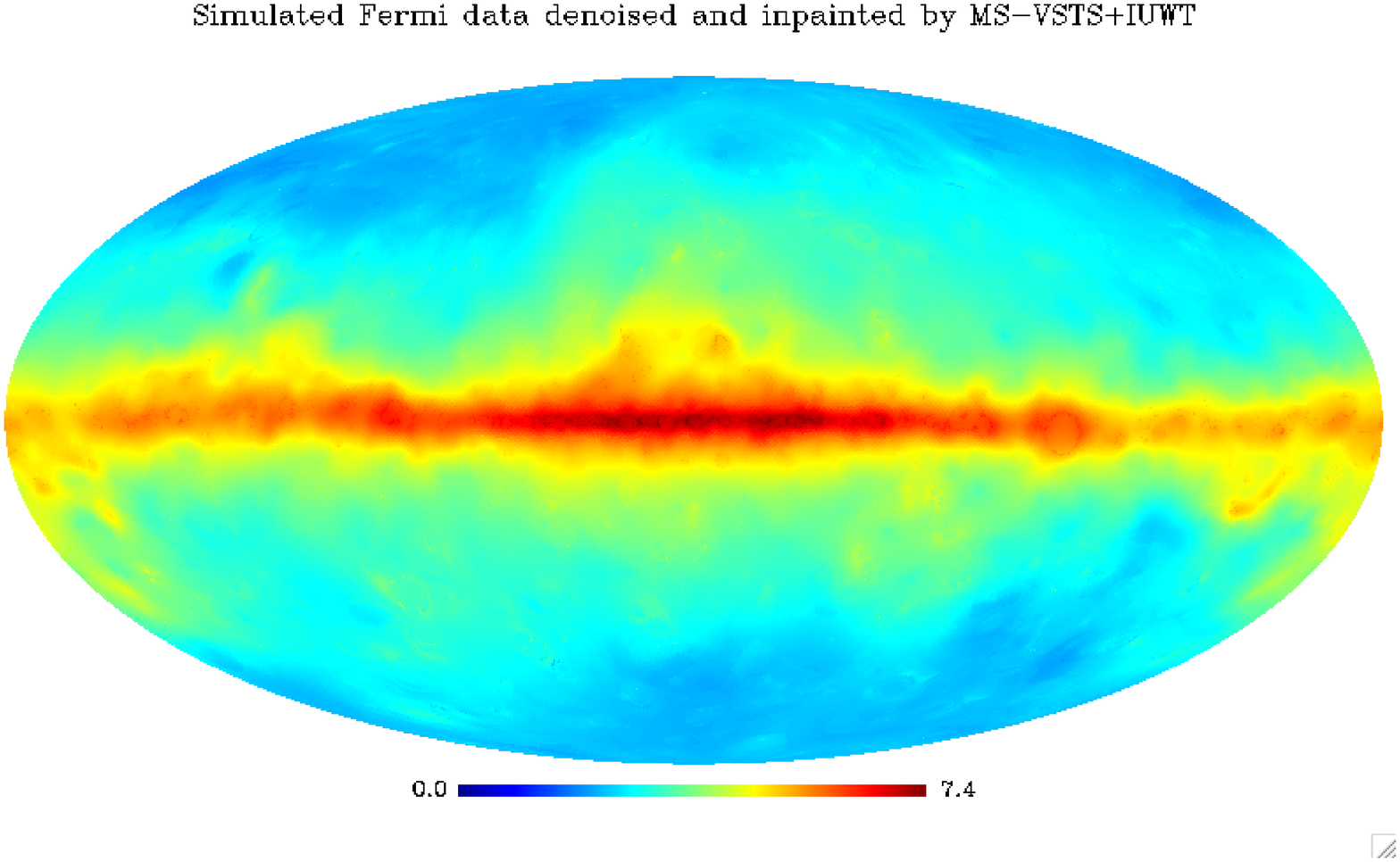} \hfill
\caption{MS-VSTS - Inpainting.
\emph{Left}: Fermi simulated map with Poisson noise and the most luminous sources masked.
\emph{Right}: Fermi simulated map denoised and inpainted with wavelets (Algorithm~\ref{alg2}).
Pictures are in logarithmic scale.
}
\label{impainting}
\end{center}
\end{figure*}

The results are on Figure~\ref{impainting}. The MS-VST + IUWT + Inpainting method (Algorithm~\ref{alg2}) interpolates the missing data very well. Indeed, the missing part can not be seen anymore in the inpainted map, which shows that the diffuse emission component  has been \textbf{correctly} reconstructed.

\section{Source detection: denoising and background modeling}
\subsection{Method}
In the case of Fermi data, the diffuse gamma-ray emission from the Milky Way, due to interaction between cosmic rays and interstellar gas and radiation, makes a relatively intense background. We have to extract this background in order to detect point sources. This diffuse interstellar emission can be modelled by a linear combination of gas templates and inverse compton map. We can use such a background model and incorporate a background removal in our denoising algorithm.

We note $\mathbf{Y}$ the data, $\mathbf{B}$ the background we want to remove, and $d^{(b)}_{j}[k]$ the MS-VSTS coefficients of $\mathbf{B}$ at scale $j$ and position $k$. We determine the multi-resolution support by comparing $|d_j[k]-d^{(b)}_{j}[k]|$ with $\kappa \sigma_j$.

We formulate the reconstruction problem as a convex constrained minimization problem:

\begin{equation}
\label{eq34}
\begin{split}
\text{Arg} \min_{\mathbf{X}} \| \mathbf{ \Phi}^{T}\mathbf{X}\|_1,
\text{s.t.} \\ \: \left\{\begin{array}{c}\mathbf{X} \geqslant 0 , \\\forall (j,k)\in \mathcal{M},      (\mathbf{ \Phi}^{T}\mathbf{X})_j[k]=(\mathbf{ \Phi}^{T}(\mathbf{Y} - \mathbf{B}))_j[k] , \end{array}\right.
\end{split}
\end{equation}

Then, the reconstruction algorithm scheme becomes:
\begin{eqnarray}
\tilde{\mathbf{X}} = P_{+}[\mathbf{ X}^{(n)} + \mathbf{ \Phi} P_{\mathcal{M}} \mathbf{ \Phi}^{T} (\mathbf{ Y} - \mathbf{B} - \mathbf{ X}^{(n)})] , \\
\mathbf{X}^{(n+1)} = \mathbf{ \Phi}\text{ST}_{\lambda_n}[\mathbf{ \Phi}^{T}\tilde{\mathbf{X}}].
\end{eqnarray}

\begin{figure*}
\begin{center}
\includegraphics[width=2.9in]{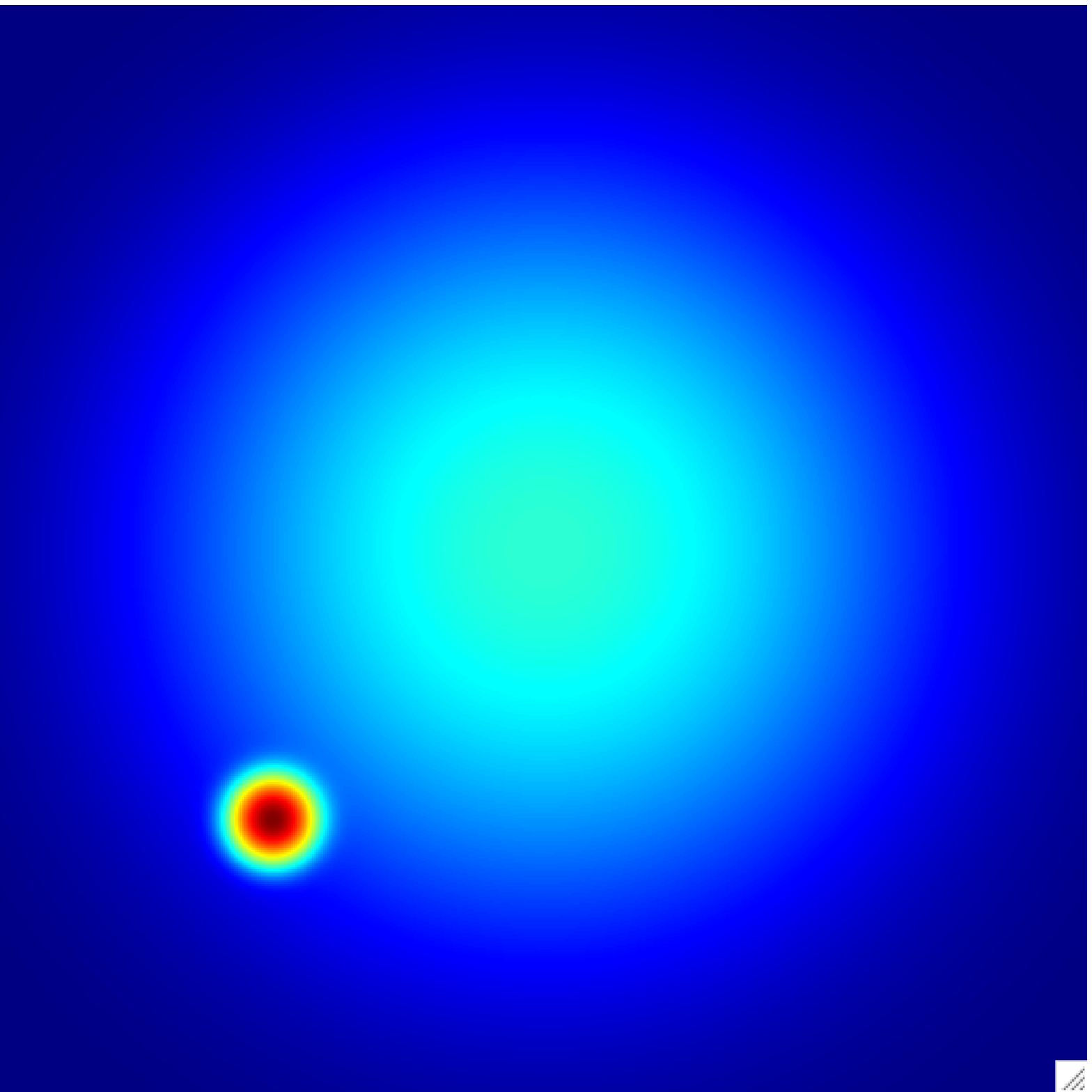} \hfill
\includegraphics[width=2.9in]{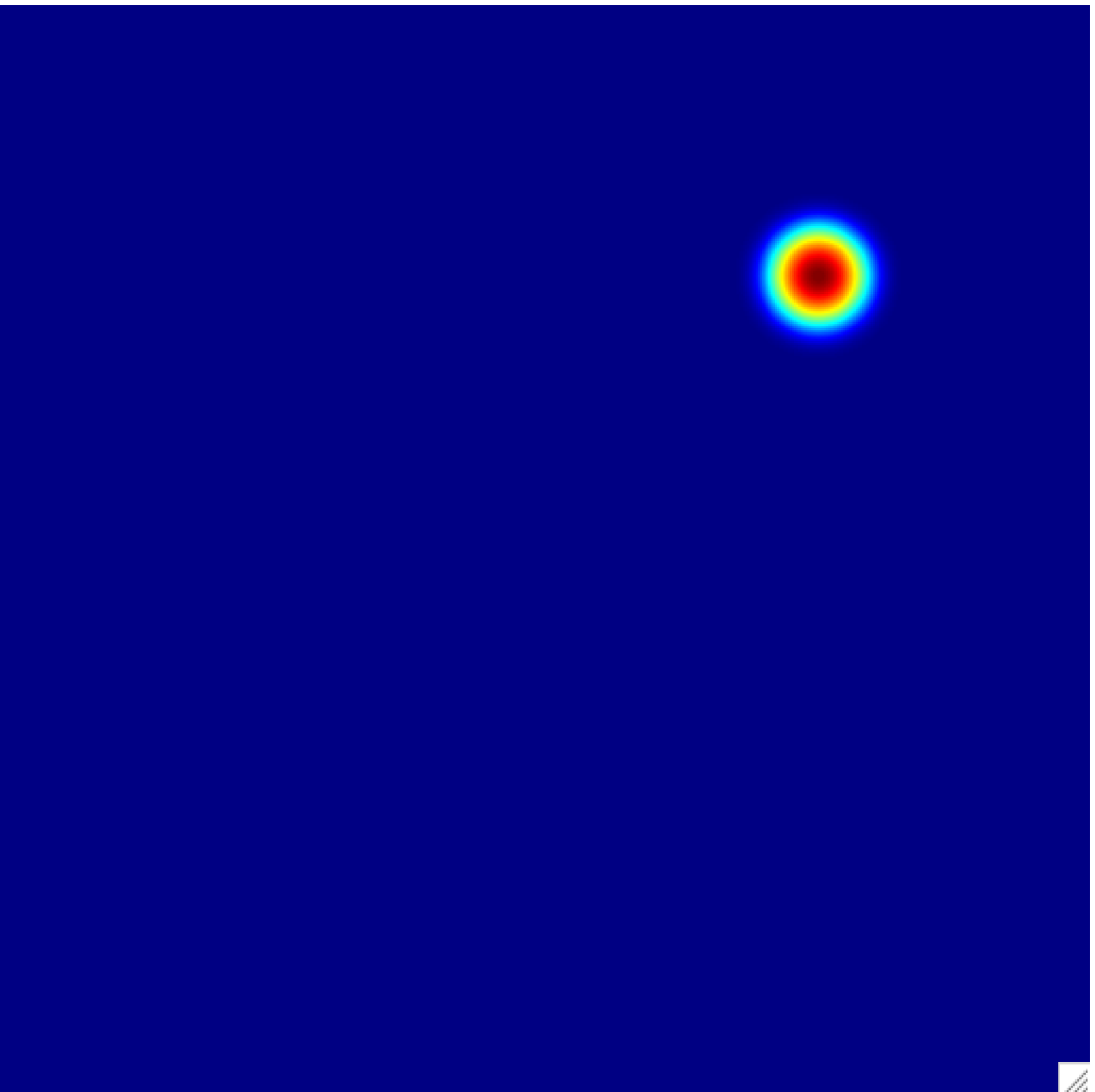} \hfill
\includegraphics[width=2.9in]{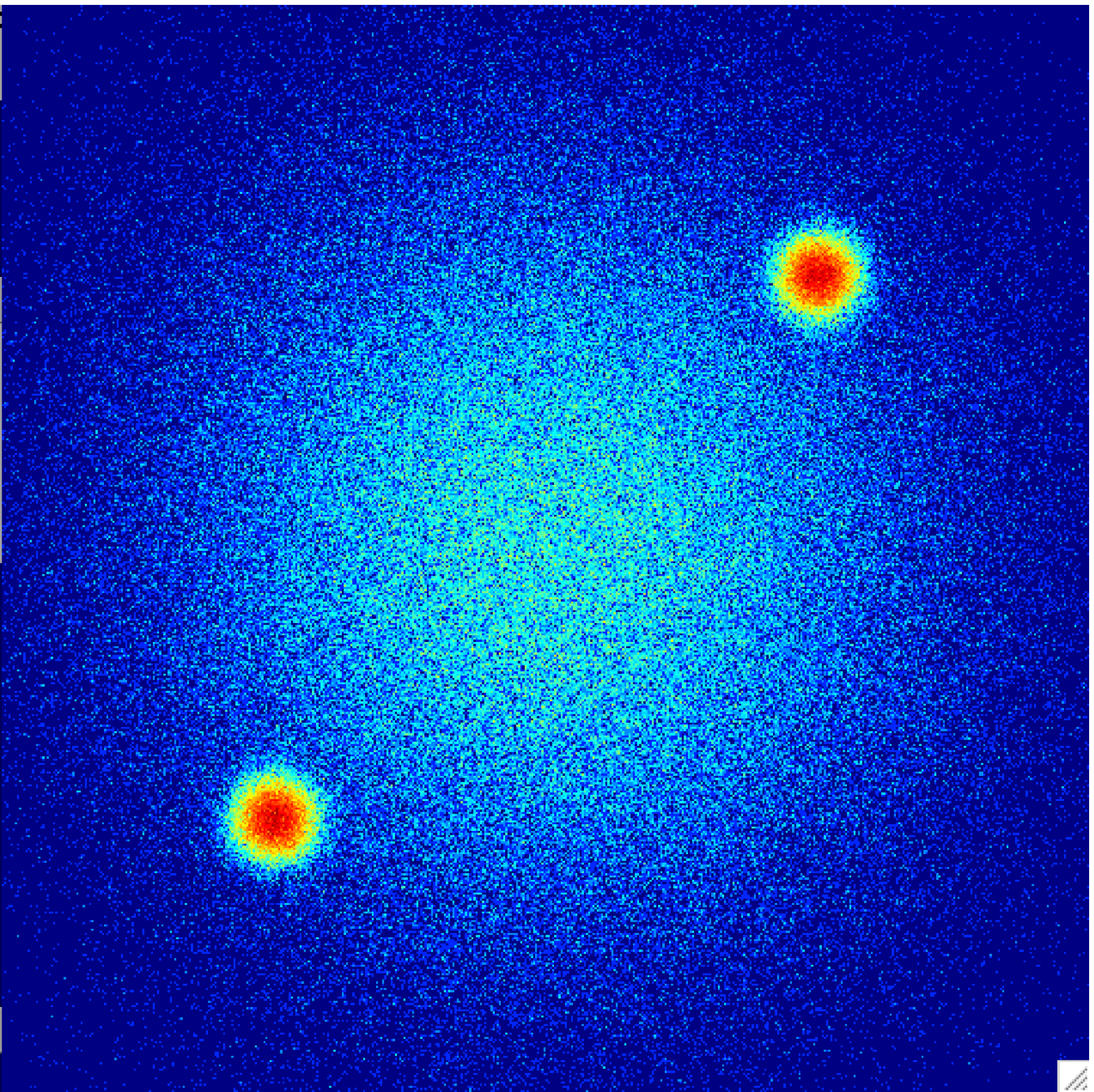} \hfill
\includegraphics[width=2.9in]{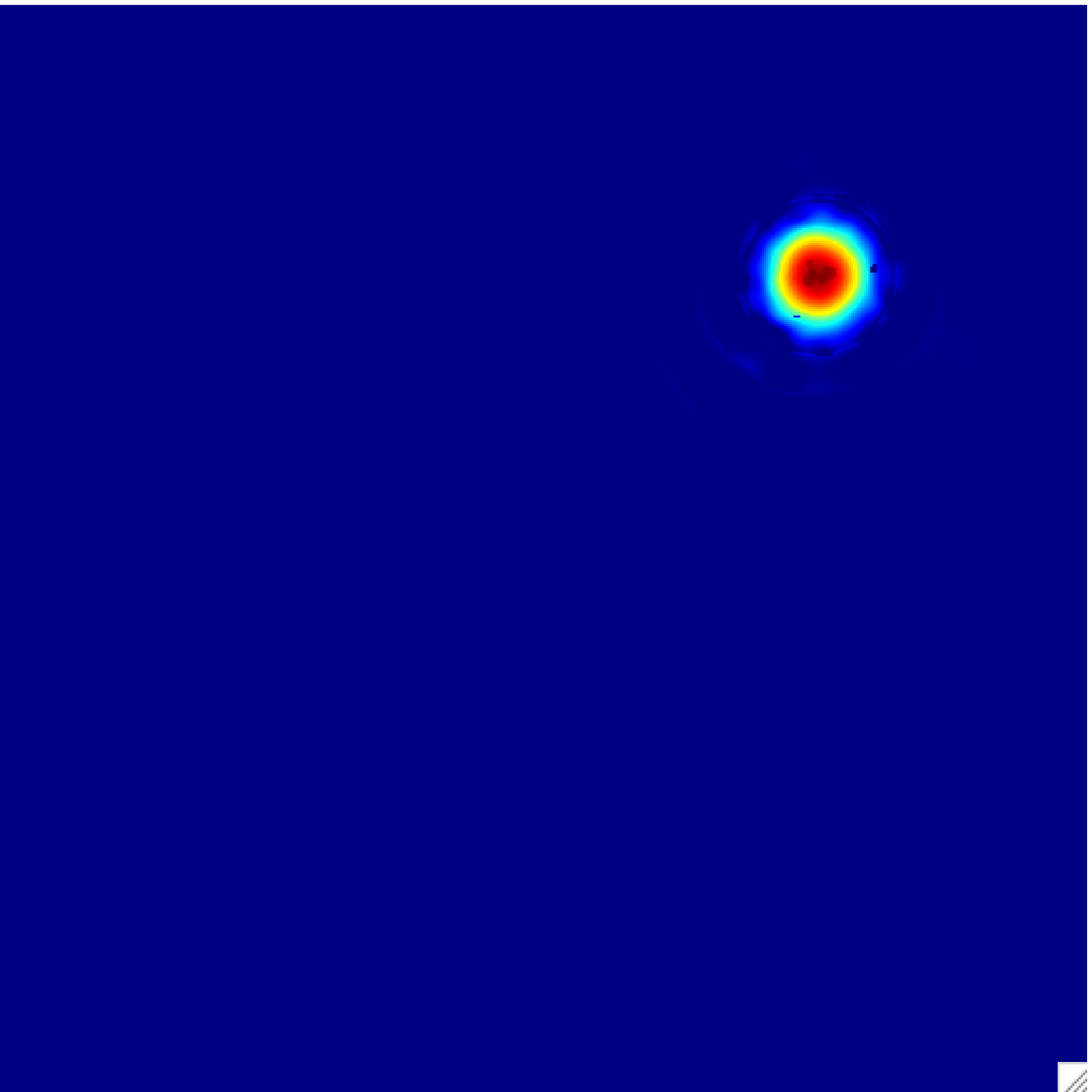}
\caption{Theoretical testing for MS-VSTS + IUWT denoising + background removal algorithm (Algorithm~\ref{alg3}). View on a single HEALPix face.
\emph{Top Left}: Simulated background : sum of two Gaussians of standard deviation equal to 0.1 and 0.01 respectively.
\emph{Top Right}: Simulated source: Gaussian of standard deviation equal to 0.01.
\emph{Bottom Left}: Simulated poisson data.
\emph{Bottom Right}: Image denoised with MS-VSTS + IUWT and background removal.
}
\label{background}
\end{center}
\end{figure*}

The algorithm is illustrated by the theoretical study \textbf{in} Figure~\ref{background}. We denoise Poisson data while separating a single source, which is a Gaussian of standard deviation equal to 0.01, from a background, which is a sum of two Gaussians of standard deviation equal to 0.1 and 0.01 respectively.

\begin{algorithm}
\caption{MS-VSTS + IUWT Denoising + Background extraction}
\label{alg3}

\begin{algorithmic}[1]
\REQUIRE $\quad$ data $a_0:=\mathbf{Y}$, background $B$, number of iterations $N_{\max}$, threshold $\kappa$. \\
\underline{\emph{\textbf{Detection}}} \\
\FOR{$j=1$ to $J$}
\STATE Compute $a_j$ and $d_j$ using (\ref{eq27}).
\STATE Hard threshold $(d_j[k] - d^{(b)}_{j}[k])$ with threshold $\kappa \sigma_j$ and update $\mathcal{M}$.
\ENDFOR \\
\underline{\emph{\textbf{Estimation}}} \\
\STATE Initialize $\mathbf{X}^{(0)}=0$, $\lambda_0 = 1$.
\FOR{$n=0$ to $N_{\max}-1$}
\STATE $\tilde{\mathbf{X}}= P_{+}[\mathbf{ X}^{(n)} + \mathbf{ \Phi} P_{\mathcal{M}} \mathbf{ \Phi}^{T} (\mathbf{ Y} - \mathbf{B} - \mathbf{ X}^{(n)})]$.
\STATE $\mathbf{X}^{(n+1)} = \mathbf{ \Phi}\text{ST}_{\lambda_n}[\mathbf{ \Phi}^{T}\tilde{\mathbf{X}}]$.
\STATE $\lambda_{n+1} = \frac{N_{\max} - (n+1)}{N_{\max} - 1}$.
\ENDFOR
\STATE Get the estimate $\hat{\mathbf{\Lambda}} = \mathbf{X}^{(N_{\max})}$.

\end{algorithmic}
\end{algorithm}

Like Algorithm~\ref{alg1}, Algorithm~\ref{alg3} can be adapted to make multiresolution support adaptation.

\subsection{Experiment}

\begin{figure*}
\begin{center}
\includegraphics[width=2.9in]{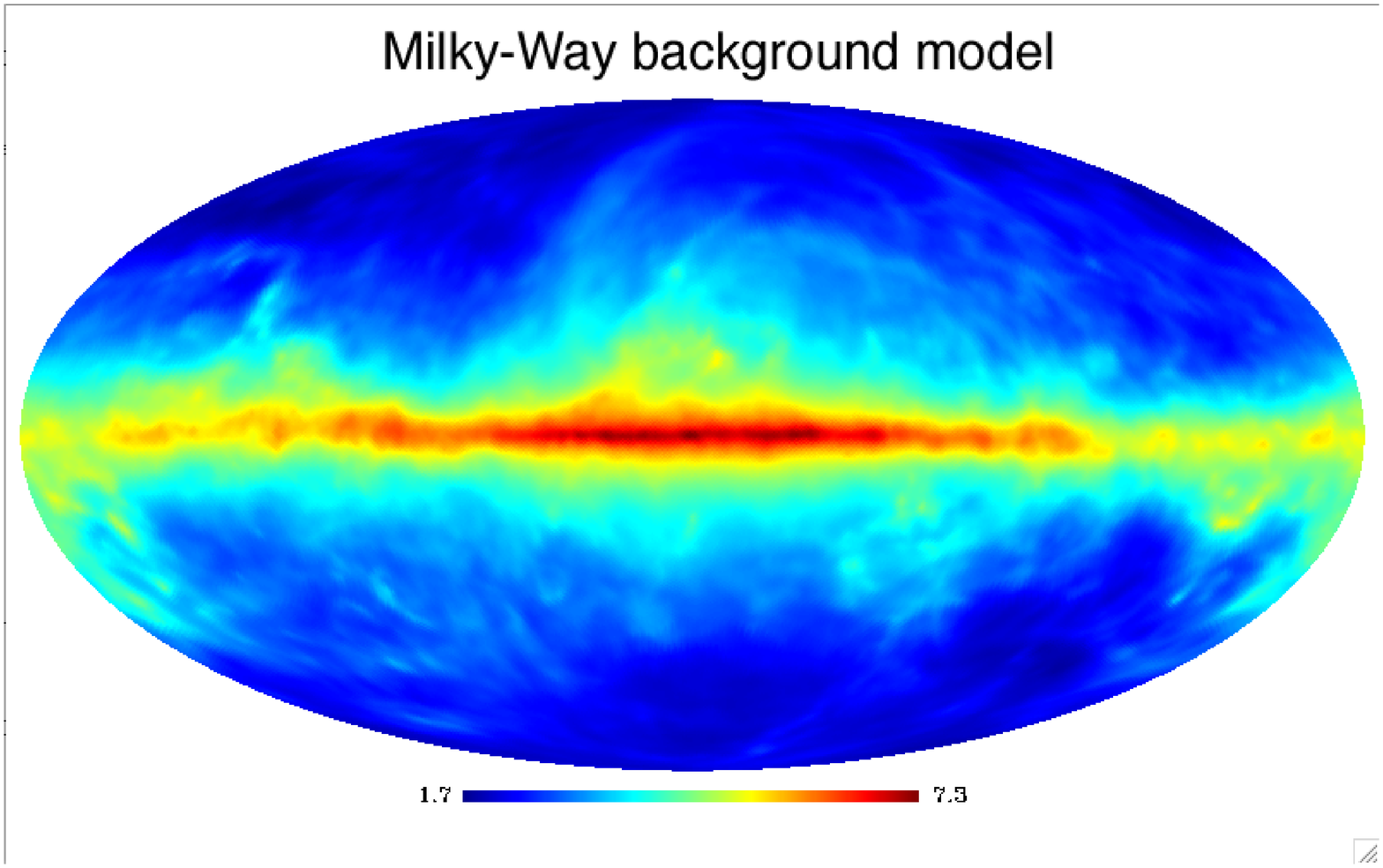} \hfill
\includegraphics[width=2.9in]{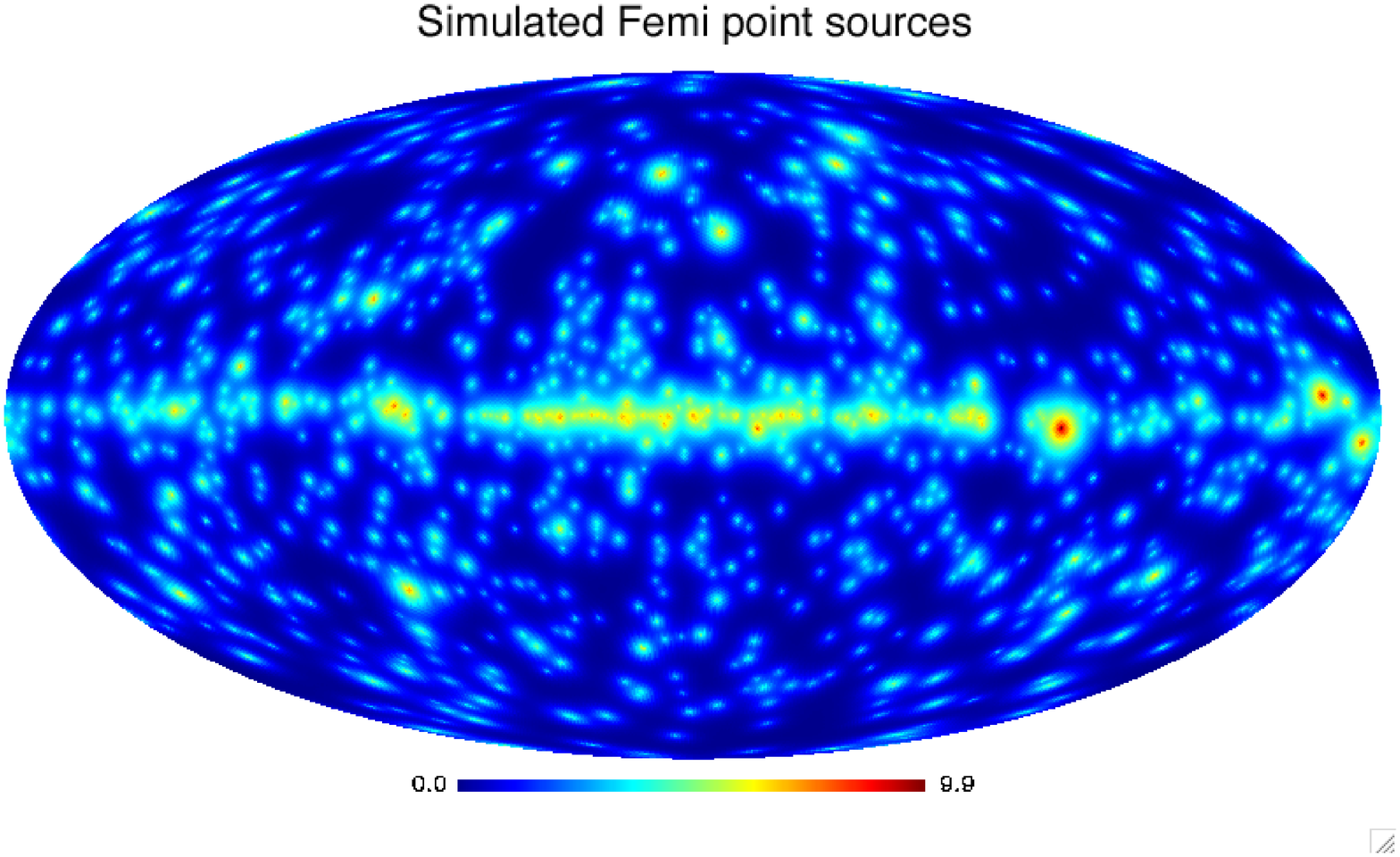} \hfill
\includegraphics[width=2.9in]{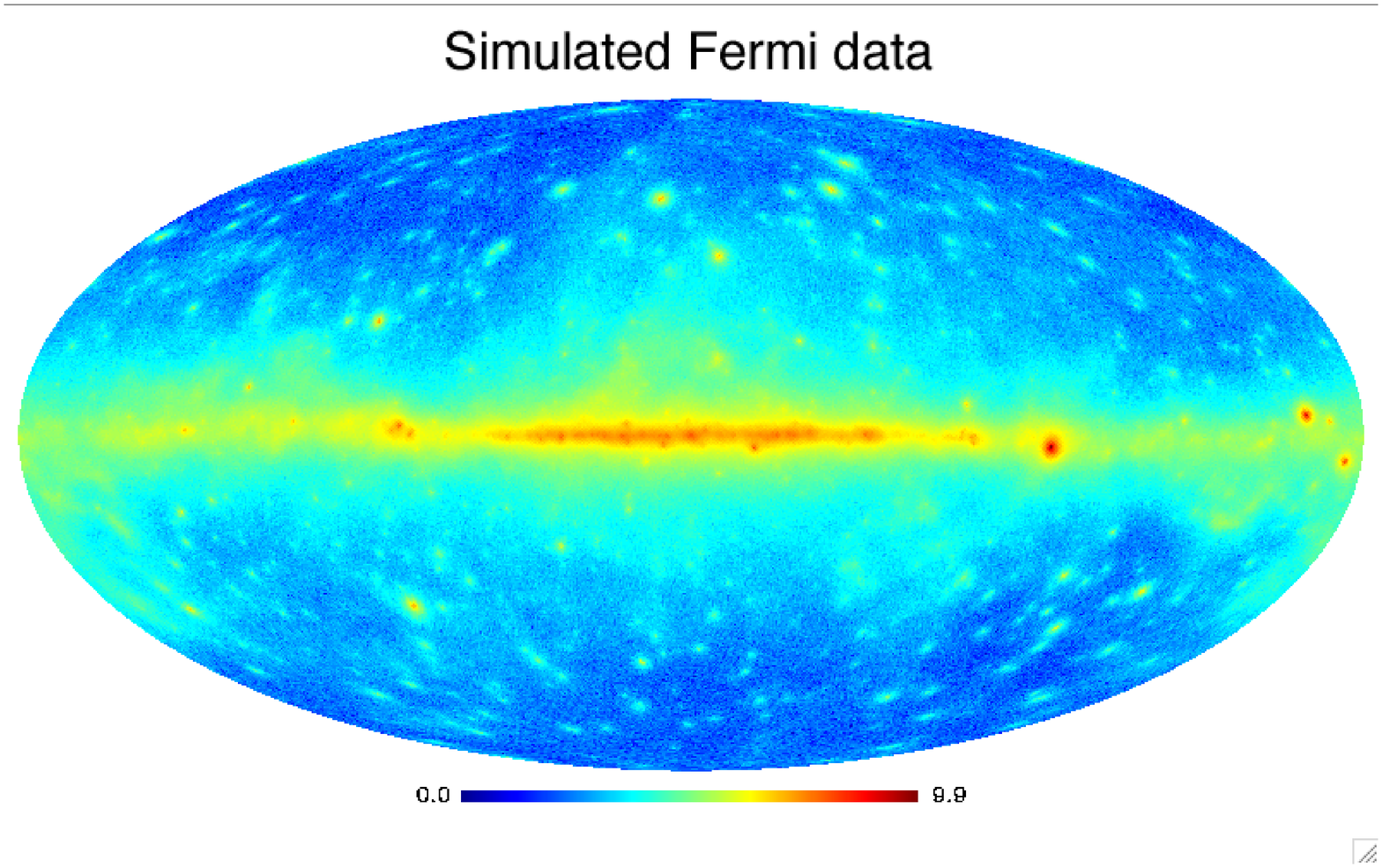} \hfill
\includegraphics[width=2.9in]{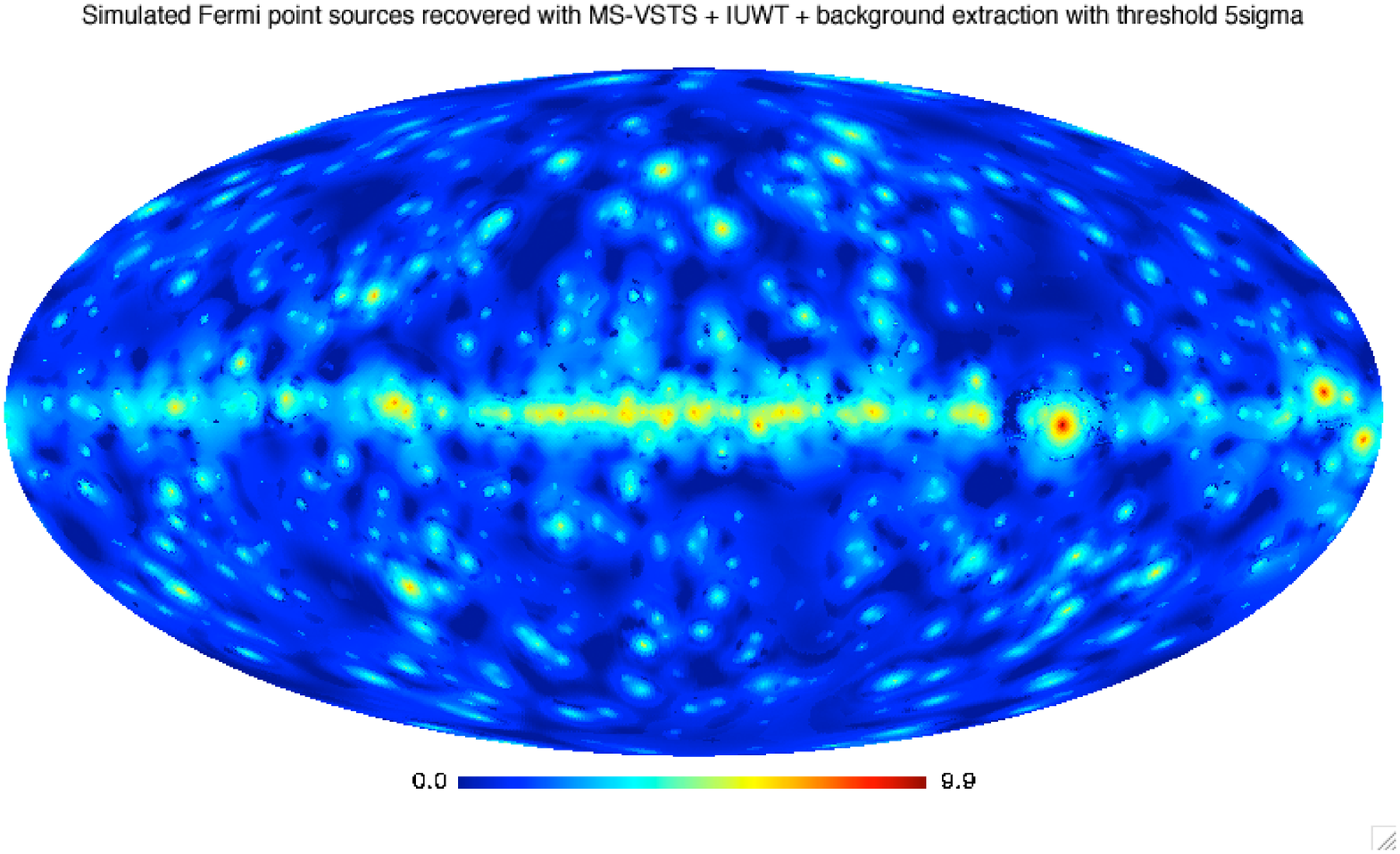} \hfill
\includegraphics[width=2.9in]{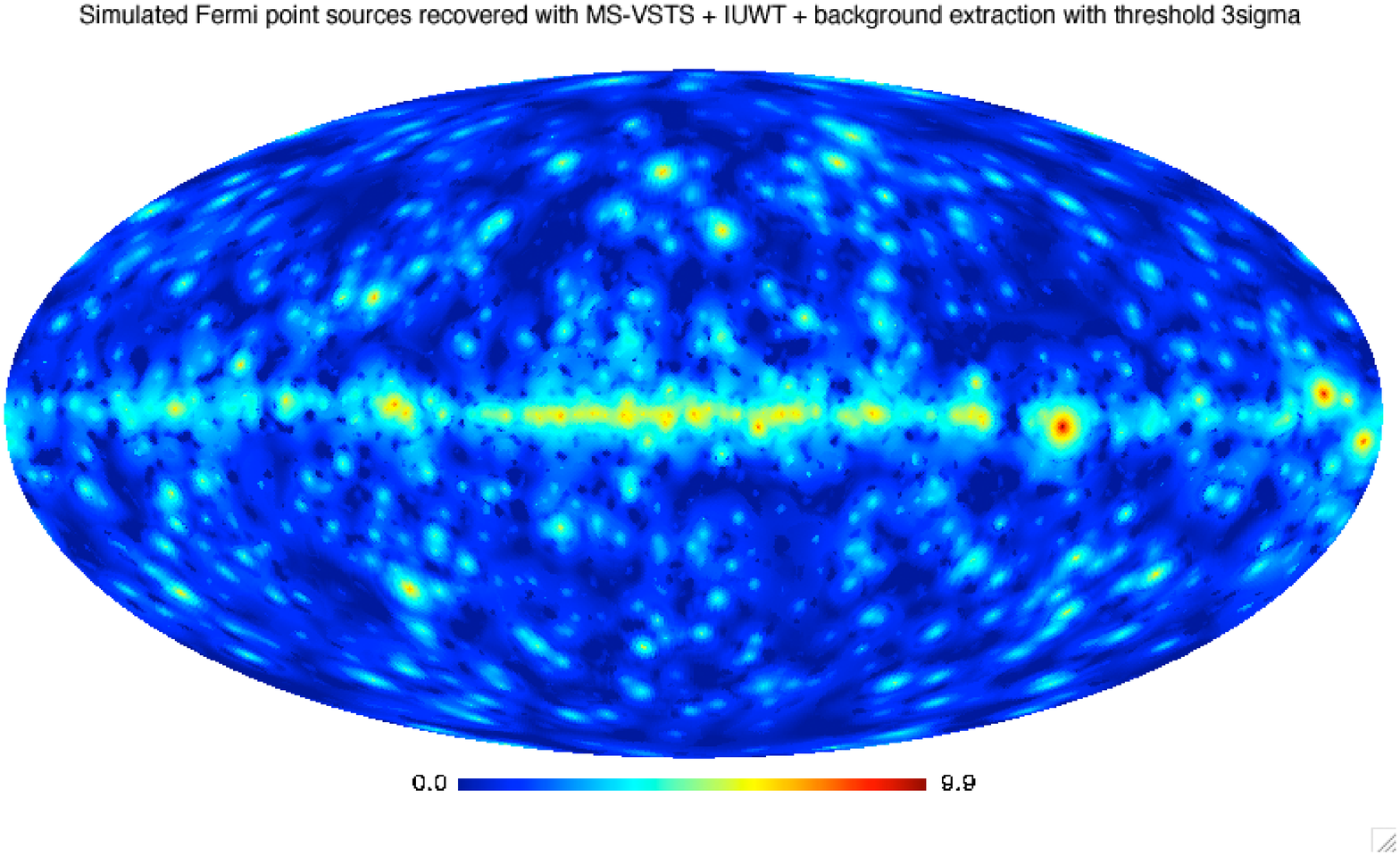}
\caption{\emph{Top Left}: Simulated background model.
\emph{Top Right}: Simulated Gamma Ray sources.
\emph{Middle Left}: Simulated Fermi data with Poisson noise.
\emph{Middle Right}: Reconstructed Gamma Ray Sources with MS-VSTS + IUWT + background removal (Algorithm~\ref{alg3}) with threshold $5\sigma_j$.
\emph{Bottom}: Reconstructed Gamma Ray Sources with MS-VSTS + IUWT + background removal (Algorithm~\ref{alg3}) with threshold $3\sigma_j$.
Pictures are in logarithmic scale.}
\label{sources}
\end{center}
\end{figure*}

\begin{figure*}
\begin{center}
\includegraphics[width=2.9in]{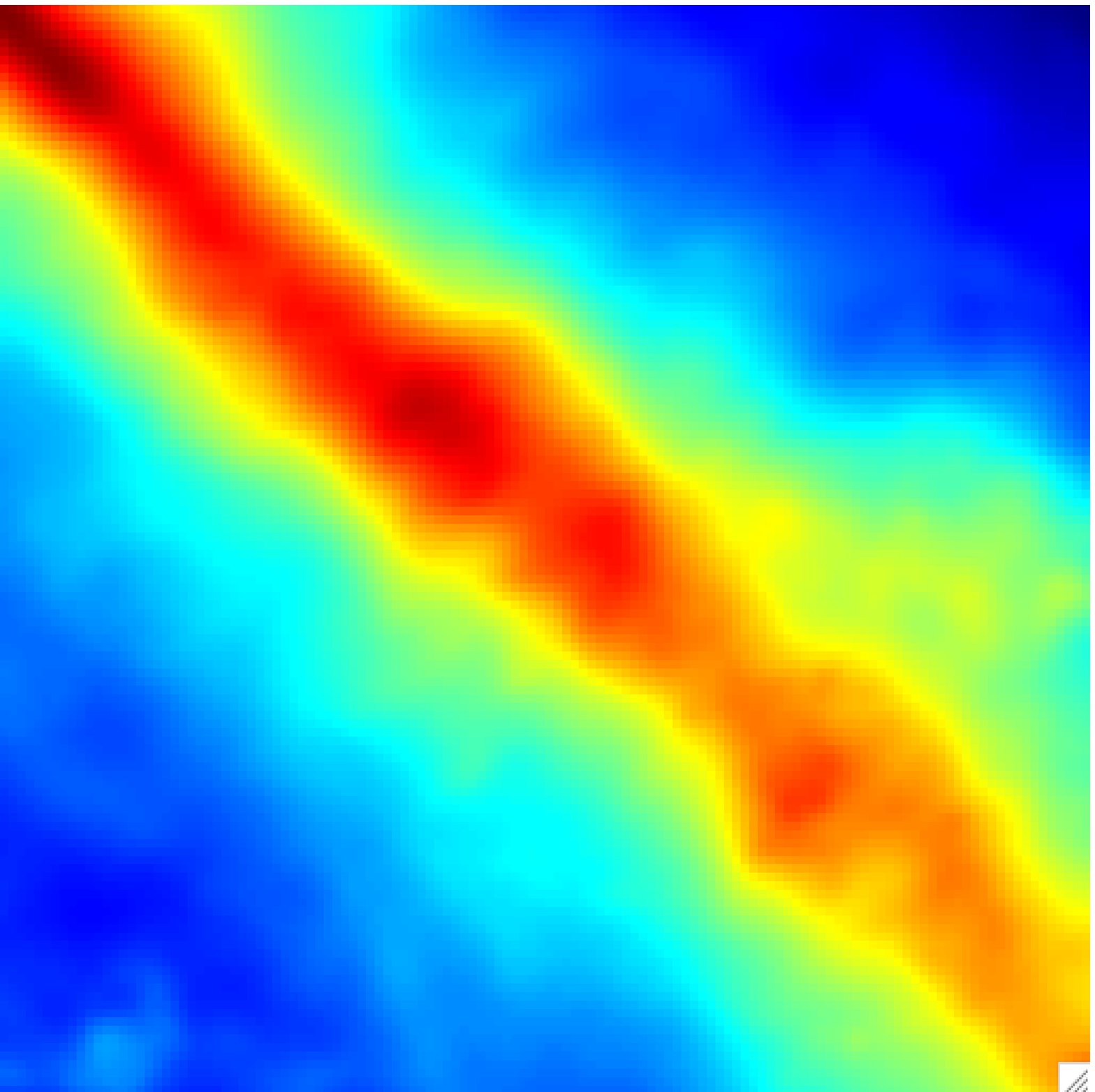} \hfill
\includegraphics[width=2.9in]{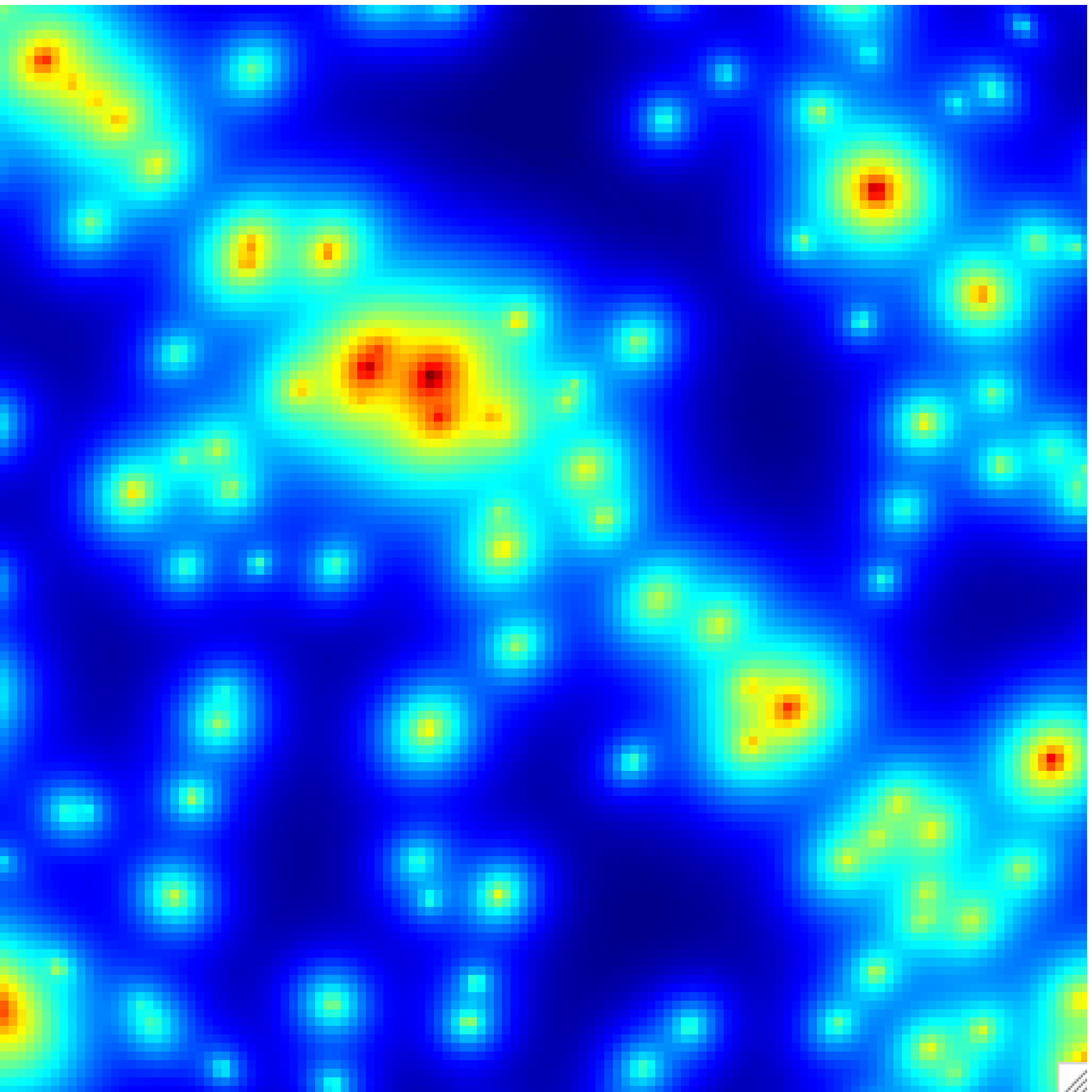} \hfill
\includegraphics[width=2.9in]{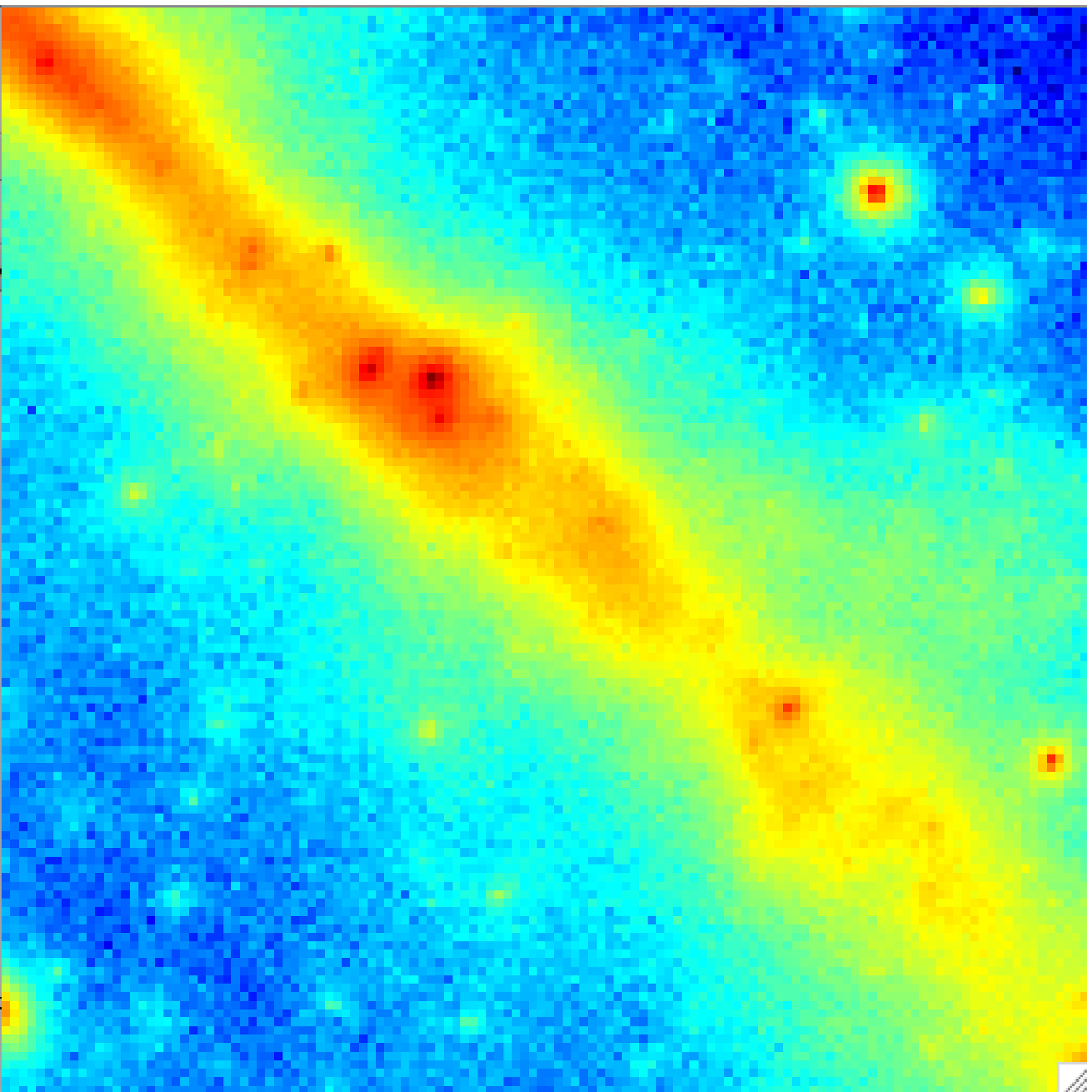} \hfill
\includegraphics[width=2.9in]{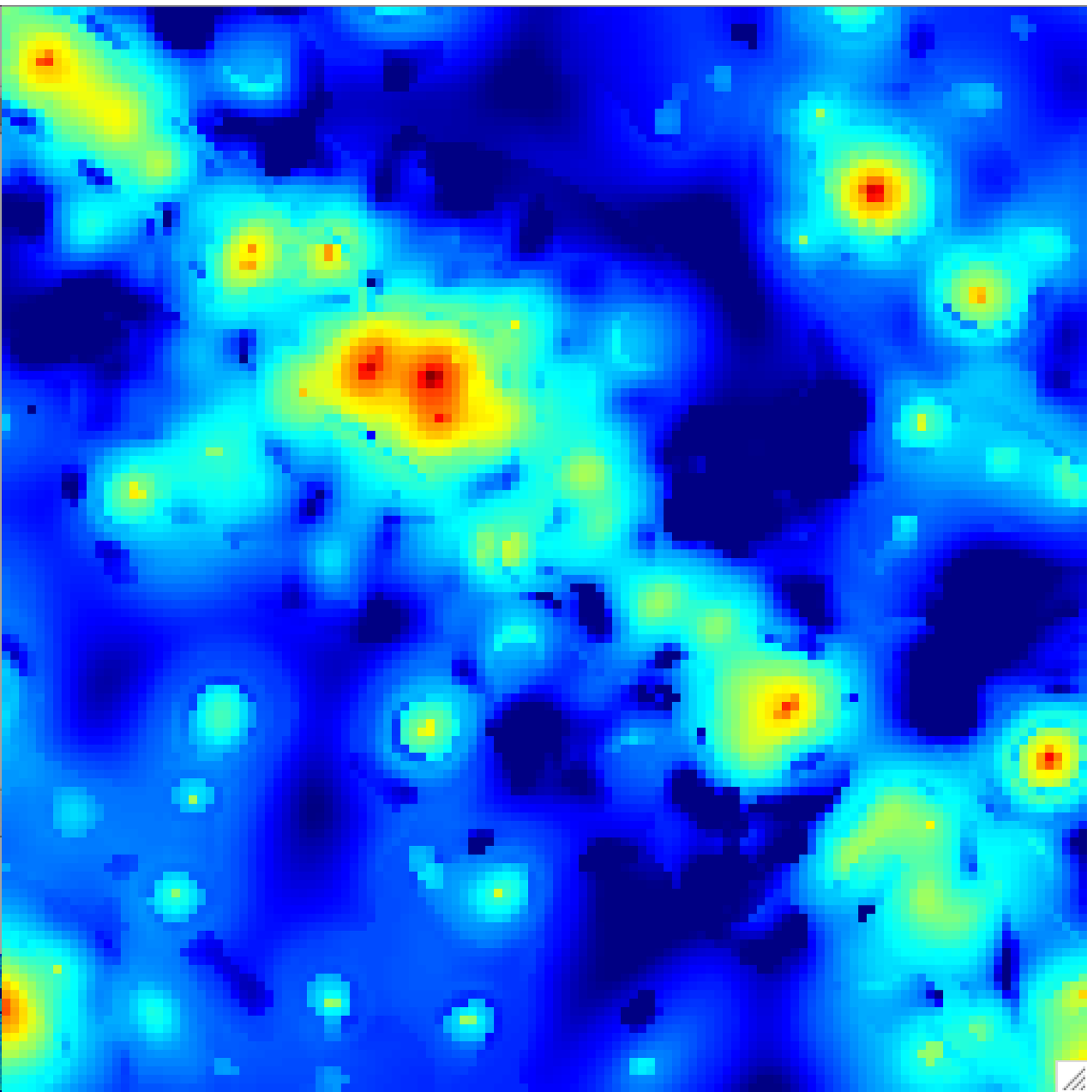} \hfill
\includegraphics[width=2.9in]{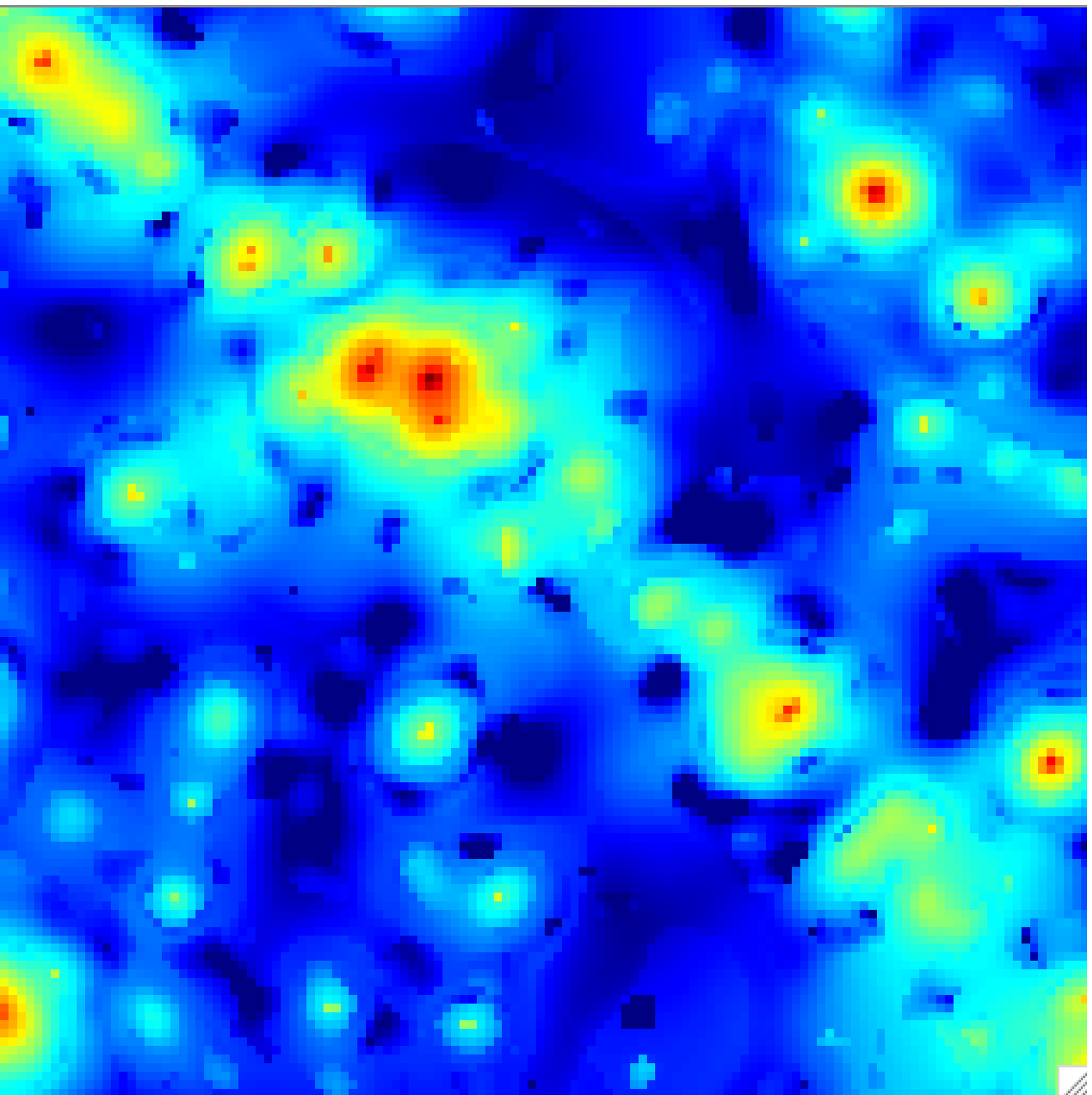}
\caption{View of a single HEALPix face from the results of Figure~\ref{sources}.\emph{Top Left}: Simulated background model.
\emph{Top Right}: Simulated Gamma Ray sources.
\emph{Middle Left}: Simulated Fermi data with Poisson noise.
\emph{Middle Right}: Reconstructed Gamma Ray Sources with MS-VSTS + IUWT + background removal (Algorithm~\ref{alg3}) with threshold $5\sigma_j$.
\emph{Bottom}: Reconstructed Gamma Ray Sources with MS-VSTS + IUWT + background removal (Algorithm~\ref{alg3}) with threshold $3\sigma_j$.
Pictures are in logarithmic scale.
}
\label{sourcesreest}
\end{center}
\end{figure*}

We applied Algorithms~\ref{alg3} on simulated Fermi data. To test the efficiency of our method, we detect the sources with the SExtractor routine~\citep{sext}, and compare the detected sources with the theoretical sources catalog to get the number of true and false detections. Results are shown on Figures~\ref{sources} and~\ref{sourcesreest}. The SExtractor method was applied on the first wavelet scale of the reconstructed map, with a detection threshold equal to 1. It has been chosen to optimise the number of true detections. SExtractor makes $593$ true detections and $71$ false detections on the Fermi simulated map restored with Algorithm~\ref{alg4} among the $1000$ sources of the simulation. On noisy data, many fluctuations due to Poisson noise are detected as sources by SExtractor, which leads to a big number of false detections (more than 2000 in the case of Fermi data).

\subsubsection{Sensitivity to model errors}

As it is difficult to model the background precisely, it is important to study the sensitivity of the method to model errors. We add a stationary Gaussian noise to the background model, we compute the MS-VSTS + IUWT with threshold $3\sigma_j$ on the simulated Fermi Poisson data with extraction of the noisy background, and we study the percent of true and false detections with respect to the total number of sources of the simulation and the signal-noise ratio ($\text{SNR} (dB) = 20 \log (\sigma_{signal} / \sigma_{noise})$) versus the standard deviation of the Gaussian perturbation. Table~\ref{table1} shows that, when the standard deviation of the noise on the background model becomes of the same range \textbf{as} the mean of the Poisson intensity distribution ($\lambda_{\text{mean}} = 68.764$), the number of false detections increases, the number of true detections decreases and the signal noise ratio decreases. While the perturbation is not too strong (standard deviation $< 10$), the effect of the model error remains low.

\begin{table*}
  \centering
  \caption{Percent of true and false detection and signal-noise ratio versus the standard deviation of the Gaussian noise on the background model. 
  }
  \begin{tabular}{|c|c|c|c|}
\hline
Model error std dev & $\%$ of true detect & $\%$ of false detect & SNR (dB) \\
\hline
  0 & $59.3\%$ & $7.1\%$ & 23.8 \\
  10 & $57.0\%$ & $11.0\%$ & 23.2 \\
  20 & $53.2\%$ & $18.9\%$ & 22.6 \\
  30 & $49.1\%$ & $43.5\%$ & 21.7 \\
  40 & $42.3\%$ & $44.3\%$ & 21.0 \\
  50 & $34.9\%$ & $39.0\%$ & 20.3 \\
  60 & $30.3\%$ & $37.5\%$ & 19.5 \\
  70 & $25.0\%$ & $34.6\%$ & 18.9 \\
  80 & $23.0\%$ & $28.5\%$ & 18.7 \\
  90 & $23.6\%$ & $27.1\%$ & 18.3 \\  
\hline
\end{tabular}
  
  \label{table1}
\end{table*}

\section{Conclusion}

This paper presented new methods for restoration of spherical data with noise following a Poisson distribution. A denoising method was proposed, which used a variance stabilization method and multiscale transforms on the sphere.
Experiments have shown it is very efficient for Fermi data denoising.  Two spherical multiscale transforms, the wavelet and the curvelets, were used.
Then, we have proposed an extension of the denoising method in order to take into account missing data, and we have shown that this inpainting method
could be a useful tool to estimate the diffuse emission. 
Finally, we have introduced a new denoising method the sphere which takes into account a background model. The simulated data have shown that it is relatively robust to
errors in the model, and can therefore be used for Fermi diffuse background modeling and source detection.

\section*{Acknowledgement}
This work was partially supported by the  European Research Council grant ERC-228261.

\bibliographystyle{aa} 
\bibliography{13822AA.bib} 
\end{document}